\documentclass[fleqn,usenatbib]{mnras}

\usepackage{newtxtext,newtxmath}

\usepackage[T1]{fontenc}

\DeclareRobustCommand{\VAN}[3]{#2}
\let\VANthebibliography\thebibliography
\def\thebibliography{\DeclareRobustCommand{\VAN}[3]{##3}\VANthebibliography}

\usepackage{graphicx}	
\usepackage{amsmath}	

\newcommand{\angstrom}{\mbox{\normalfont\AA}}

\title[The miniJPAS survey quasar selection II]{The miniJPAS survey quasar selection II: Machine Learning classification with photometric measurements and uncertainties}

\author[N.V.N. Rodrigues et al.]{
\parbox[t]{\textwidth}{
Nat\'alia V.N. Rodrigues$^{1}$\thanks{Email: natalia.villa.rodrigues@usp.br},
        L. Raul Abramo$^{1}$, 
        Carolina Queiroz$^{1,2}$,
        Gin\'es Mart\'inez-Solaeche$^{3}$,
        Ignasi P\'erez-R\`afols$^{4,5}$,
        Silvia Bonoli$^{7,8}$,
        Jon\'as Chaves-Montero$^{7}$,
        Matthew M. Pieri$^{6}$,
        Rosa M. Gonz\'alez Delgado$^{3}$,
        Sean S. Morrison$^{6,9}$,
        Valerio Marra$^{10,11}$
        Isabel M\'arquez$^{3}$,
        A. Hern\'an-Caballero$^{17}$,
        L.A. D\'iaz-Garc\'ia$^{3}$,
        Narciso Ben\'itez$^{3}$,
        A. Javier Cenarro$^{16}$,
        Renato A. Dupke$^{12,14,15}$,
        Alessandro Ederoclite$^{17}$,
        Carlos L\'opez-Sanjuan$^{16}$,
        Antonio Mar\'in-Franch$^{16}$,
        Claudia Mendes de Oliveira$^{13}$,
        Mariano Moles$^{3,17}$,
        Laerte Sodr\'e Jr.$^{13}$,
        Jes\'us Varela$^{16}$,
        H\'ector V\'azquez Rami\'o$^{16}$ and
        Keith Taylor$^{18}$}
\\
\\
\parbox[t]{\textwidth}{
$^{1}$Departamento de F\'isica Matem\'atica, Instituto de F\'{\i}sica, Universidade de S\~ao Paulo, Rua do Mat\~ao 1371, CEP 05508-090, S\~ao Paulo, Brazil\\
   $^{2}$Departamento de Astronomia, Instituto de F\'isica, Universidade Federal do Rio Grande do Sul (UFRGS), Av. Bento Gon\c{c}alves 9500, Porto Alegre, RS, Brazil\\
   $^{3}$Instituto de Astrof\'isica de Andaluc\'ia (CSIC), P.O. Box 3004, 18080 Granada, Spain\\
   $^{4}$Institut de Física d’Altes Energies (IFAE), The Barcelona Institute of Science and Technology, 08193 Bellaterra (Barcelona), Spain\\
   $^{5}$Sorbonne Universit\'e, Universit\'e Paris Diderot, CNRS/IN2P3, Laboratoire de Physique Nucl\'eaire et de Hautes Energies, LPNHE, 4 Place Jussieu, F-75252 Paris, France\\
   $^{6}$Aix Marseille Univ, CNRS, CNES, LAM, Marseille, France\\
   $^{7}$Donostia International Physics Center, Paseo Manuel de Lardizabal 4, E-20018 Donostia-San Sebastian, Spain\\
   $^{8}$Ikerbasque, Basque Foundation for Science, E-48013 Bilbao, Spain\\
   $^{9}$Department of Astronomy, University of Illinois at Urbana-Champaign, Urbana, IL 61801, USA\\
   $^{10}$ INAF, Osservatorio Astronomico di Trieste, via Tiepolo 11, 34131 Trieste, Italy\\
   $^{11}$  IFPU, Institute for Fundamental Physics of the Universe, via Beirut 2, 34151, Trieste, Italy\\
   $^{12}$ Observat\'orio Nacional/MCTI, Rua General Jos\'e Cristino, 77, São Crist\'ov\~ao, CEP 20921-400, Rio de Janeiro, Brazil\\
   $^{13}$ Universidade de S\~ao Paulo, Instituto de Astronomia, Geof\'isica e Ci\^encias Atmosf\'ericas, Depto. de Astronomia, Rua do Mat\~ao, 1226, CEP 05508-090, S\~ao Paulo, Brazil\\
   $^{14}$ Department of Astronomy, University of Michigan, 311 West Hall, 1085 South University Ave., Ann Arbor, USA\\
   $^{15}$ University of Alabama, Department of Physics and Astronomy, Gallalee Hall, Tuscaloosa, AL 35401, USA\\
   $^{16}$ Centro de Estudios de F\'isica del Cosmos de Arag\'on (CEFCA), Unidad Asociada al CSIC, Plaza San Juan 1, 44001, Teruel, Spain\\
   $^{17}$ Centro de Estudios de F\'isica del Cosmos de Arag\'on (CEFCA), Plaza San Juan, 1, 44001, Teruel, Spain\\
   $^{18}$ Instruments4, 4121 Pembury Place, La Canada Flintridge, CA 91011, USA
\\}
}

\date{Accepted 2022 September 27. Received 2022 September 26; in original form 2022 August 22}

\pubyear{2023}

\begin{document}
\label{firstpage}
\pagerange{\pageref{firstpage}--\pageref{lastpage}}
\maketitle

\begin{abstract}
Astrophysical surveys rely heavily on the classification of sources as stars, galaxies or quasars from multi-band photometry.
Surveys in narrow-band filters allow for greater discriminatory power,
but the variety of different types and redshifts of the objects present a challenge to standard template-based methods.
In this work, which is part of larger effort that aims at building a catalogue of quasars from the miniJPAS survey, we present a Machine Learning-based method that employs Convolutional Neural Networks (CNNs) to classify point-like sources including the information in the measurement errors.
We validate our methods using data from the miniJPAS survey, a proof-of-concept project of the J-PAS collaboration covering $\sim$ 1 deg$^2$ of the northern sky using the 56 narrow-band filters of the J-PAS survey.
Due to the scarcity of real data, we trained our algorithms using mocks that were purpose-built to reproduce the distributions of different types of objects that we expect to find in the miniJPAS survey, as well as the properties of the real observations in terms of signal and noise.
We compare the performance of the CNNs with other well-established Machine Learning classification methods based on decision trees, finding that the CNNs improve the classification when the measurement errors are provided as inputs.
The predicted distribution of objects in miniJPAS is consistent with the putative luminosity functions of stars, quasars and unresolved galaxies.
Our results are a proof-of-concept for the idea that the J-PAS survey will be able to detect unprecedented numbers of quasars with high confidence.
\end{abstract}




\section{Introduction}

Galaxy surveys 
have evolved to tackle a broad range of fundamental questions, from dark energy and neutrino masses to galaxy evolution and the halo-galaxy connection  \citep{2dF,SDSS2021,WIGGLEZ,HSC,DESCosmo2021}.
Technological advances and investment in new instruments have amplified the scope of these surveys, which demand increasingly sophisticated toolboxes for data reduction, statistical analysis and phenomenology.

The first step in any survey is finding luminous sources behind the foregrounds of the sky and the Milky Way -- a task which is often performed using optical data. 
Typically, a large number of sources is detected using photometry in broad optical filters, but only a small fraction of those sources are then selected for spectroscopic follow-up observations. 
This target selection can be made on the basis of the multi-band photometry, by inspecting variability in the time-domain \citep{Morganson2015ApJ,LSST_ivezic}, by cross-matching the sources with other wavelengths \citep{WISE,XMM_xray}, or by some combination thereof.
In fact, the decision process about which of those luminous sources are likely to be the kinds of objects of interest to a given survey is the crucial first step which determines how we employ valuable resources, such as a multi-object spectrograph on a large telescope. 

The \textit{Javalambre Physics of the Accelerating Universe Astrophysical Survey} (J-PAS, \citet{benitez2014jpas}) was designed to take multi-band photometry in narrow filters (of width $\sim 100$ \AA) of all sources in its field of view, providing low-resolution spectra ($R\sim60$) 
in the interval $3,500 \AA \lesssim \lambda \lesssim 9,000 \AA$
-- in that context, see also \citet{Combo17,PAUS} for other narrow-band surveys.
The {\em science verification} phase of the survey, {\em miniJPAS} \citep{miniJPAS}, achieved 5-$\sigma$ limiting magnitudes (for an aperture of $3''$) of approximately $\sim 23-24$ for the broad bands ($u$, $g$, $r$ and $i$), and between $\sim 22-23$ for the narrow bands.
MiniJPAS has demonstrated that optical ``pseudo-spectra'' are often sufficient to determine with high confidence whether an object is a star, a galaxy, a quasar or some other type of source -- and, in the case of extragalactic sources, to determine the redshifts of those objects with sub-percent precision.

However, even with exquisite photometry a precise determination of the classes of very large numbers (millions, or even billions) of objects is a challenge to established methods such as magnitude and/or color cuts, as well as techniques that rely on template fitting \citep{Takada2014PASJ,Dawson2016AJ}. 
This is particularly problematic in the case of rare objects such as quasars, which can be drowned by the heaps of stars and galaxies that constitute the bulk of sources in photometric surveys \citep{Myers2015ApJS,Dwelly2017MNRAS}.

Given the advantages of narrow-band photometry to classify astrophysical sources, and in particular objects with strong emission lines such as quasars \citep{eldar}, the J-PAS and WEAVE-QSO \citep{pieri2016weaveqso} surveys have partnered to produce the largest, most complete high-redshift quasar survey to-date.
The goal is to build a near-complete sample of quasars identified with the help of the J-PAS multi-band photometry (hereafter J-spectra), targeting in particular the $z\geq2.1$ quasars for follow-up using the WEAVE multi-object spectrograph \citep{dalton2016weave}. 
The WEAVE instrument will confirm whether those objects are really quasars, helping refine the J-PAS classification and redshift estimates. 
WEAVE will also be able to measure the Ly-$\alpha$ absorption systems along the lines of sight to those high-redshift quasars, providing crucial information about the large-scale structures along those lines of sight. 
This data set, which will eventually cover approximately 6000 deg$^2$, will allow us to compute the clustering of matter using both the Ly-$\alpha$ systems and the quasars themselves, measuring distances using the baryon acoustic oscillation scale and imposing constraints on cosmological parameters at high redshifts.

In this paper we show how Machine Learning (ML) techniques can be used to classify astrophysical objects using as input data the J-spectra yielded by multi-band photometric data, including the measurement errors.
Here we employ only photometric features such as the fluxes and their associated errors.
Additional features, such as morphology, time-domain or other ancillary data, were not included in our analysis at this moment.

The main innovation in this paper is a systematic inclusion of information about the uncertainties in the fluxes, which are key ingredients of any measurements, but are often ignored in ML applications that take scientific data as input \citep{Reis_2018, pbaqui, 2020arXiv200110523V, Shy_2022}.
Here we focus on Convolutional Neural Networks (CNNs, \citealt{CNN_lecun}), which have been developed primarily as tools to extract features from 2D images \citep{2014arXiv1409.1556S}.
CNNs have also been employed for classification purposes in astrophysics due to its general ability to detect features in images \citep{2019MNRAS.490.3952B,2019A&A...621A..26P}, on multi-band photometric data \citep{2020MNRAS.491.2280S}, and even in the time-spectral domain \citep{scone}.
It is straightforward to apply CNNs to sequential data, and to incorporate the information about measurement errors -- for a general description of the technique see also \citet{rodrigues2021information}.
In order to compare our CNN-based techniques with other well-established ML classification methods we have also tested the performance of  Random Forests \citep{breiman2001rf} and the Light Gradient Boosting Machine (LightGBM, \citealt{lgbm}), two powerful Decision Tree (DT, \citealt{DT}) based algorithms.

This work is part of a larger effort to classify miniJPAS point-like sources.
The first paper \citep{queiroz2022minijpas} describes the construction of simulated data sets ({\em mocks}) that we used to train our algorithms, and in this paper we apply CNN and DT-based ML models to those mocks.
In particular, we present a technique that enables us to take into account the measurement errors in the J-spectra.
We evaluate the performances of the classifiers not only with respect to validation data sets, but also for the real miniJPAS point sources, by comparing the numbers of objects with those expected from the luminosity functions in different magnitude ranges, redshift ranges and for the different stellar types.
Finally, we test the robustness of the classification against changes in the training sets, and we perform a feature importance analysis to evaluate which miniJPAS filters are more relevant to distinguish between the different classes.

In a closely related work, Martínez-Solaeche et al. (in preparation) focuses on a class of well-established ML models, the artificial neural networks, to explore different input features as well as to implement data augmentation techniques that introduce hybrid objects (ad-mixtures of single, pure populations) and study how this affects the confidence of the classification.
In another forthcoming paper (Pérez-R\`afols et al., in preparation), a spectral fitting method \citep{squeze} is used to estimate the probability that an object is a quasar at a given redshift.
Finally, in Pérez-R\`afols et al. (in preparation), we will show how to combine all the previous classifiers, as well as any additional external information, into a ``consensus'' catalog of stars, galaxies, low-redshift ($z < 2.1$) and high-redshift ($z \geq 2.1$) quasars.
That combined classification will constitute the final output of our mocks and of our suite of ML techniques, and will be validated with the help of the spectroscopically confirmed miniJPAS sources (the ``truth table'').

The paper is organized as follows. 
In \S\ref{Data} we describe the real and the mock data sets.
In \S\ref{ML} we introduce the methodology and the ML algorithms. 
In \S\ref{results mocks} we evaluate the performance of the models and present the results when the methods are applied to the mock test sets. 
In \S\ref{results miniJPAS} we show the results for the point sources in the miniJPAS data. 
Finally, in \S\ref{conclusion} we draw our main conclusions, and give perspectives for future improvements and applications.

\section{Data}\label{Data}
In this Section we describe the miniJPAS data sample, and briefly introduce the mocks used to train and validate the ML models -- a full description of the method used in the construction of the mocks, as well as tests used to compare them to the miniJPAS data, can be found in \citet{queiroz2022minijpas}.

\subsection{The J-PAS and miniJPAS surveys}\label{Data miniJPAS}
J-PAS is soon starting full survey operations, using a 1.2-Gpixel camera mounted on a telescope with a 2.55 m mirror and a field of view of 4.2 deg$^2$ \citep{benitez2014jpas}. 
The J-PAS photometric system \citep{2012SPIE.8450E..3SM} consists of 54 narrow-band filters and two medium-band filters (named uJAVA and J1007).
In 2020, before the full instrument was completed, the J-PAS Pathfinder camera conducted a $\sim$ 1 deg$^2$ science verification survey (the miniJPAS survey) on the area of the All-wavelength Extended Groth Strip International Survey (AEGIS, \citet{Davis_2007_aegis}). In addition to the narrow-band and medium-band filters, miniJPAS includes four SDSS-like filters $u, g, r$ and $i$ (total of 60 filters).
The primary catalogue contains 64,293 sources, and is estimated to be complete for point sources up to a magnitude of $r \simeq 23.6$. 
More details about miniJPAS can be found in \citet{miniJPAS}.

Starting from the dual mode photometry catalogue we make a quality cut that eliminates all objects with any of the flags that could indicate a problem with the photometry in any of the filters. 
This first cut lowers the number of sources down to 46,440 objects.
Next, since we are not interested in extended sources (these are almost unequivocally classified as galaxies), we selected only the point-like sources from the miniJPAS full sample, by imposing the cut $\texttt{ERT} \geq 0.1$, which is a stellarity index constructed from image morphological information, with the help of Extremely Randomized Trees \citep{pbaqui}, and which is provided in the miniJPAS catalogue. 
If that classification failed ($\texttt{ERT} = -99.0$), we then used the stellar-galaxy locus classification, with a cut of $\texttt{SGLC} \geq 0.1$ \citep{SGLC}.
After these refinements we end up with 11,419 sources
that we must now classify as either stars, galaxies, low redshift ($z < 2.1$) quasars, or high-redshift ($z \geq 2.1$) quasars\footnote{The $z = 2.1$ pivot was chosen because of the Lyman-$\alpha$ feature. 
Hence, our classification provides a preliminary sample of high-redshift quasars which will be improved with appropriate redshift estimators.}.
We then extract the fluxes and flux errors for all these objects in each filter, using the photometry for a fixed aperture of $3''$ and correcting for the light profile outside of that area, as detailed in \citet{queiroz2022minijpas}.
We refer to this sample as the miniJPAS point-like sources sub-sample.

\begin{figure}
    \centering
    \includegraphics[width=1.0\linewidth]{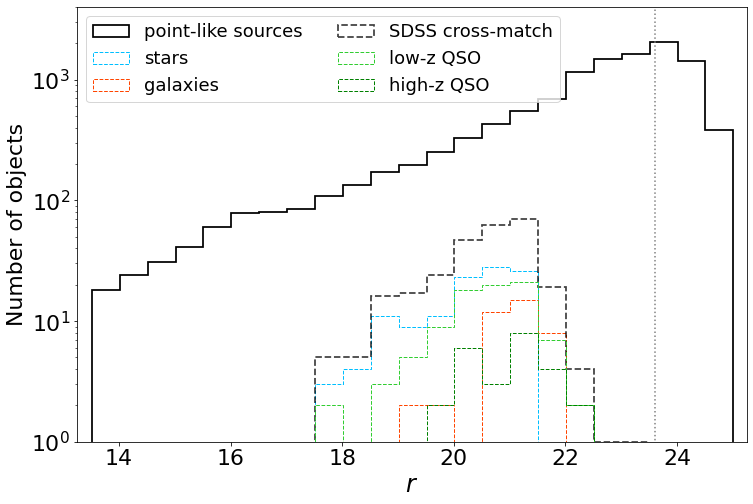}
    \caption{Histogram of the $r$ magnitudes of the miniJPAS point-like sources sub-sample (solid line), compared with those for the SDSS cross-match sample (dashed lines). The distribution of objects classified by SDSS as stars, galaxies, low-z and high-z QSO are shown in coloured dashed lines.
    The cross-match sample is effectively limited at $r \lesssim 22$, while the miniJPAS sample reachs up to $r \lesssim 24.0$. The vertical dotted line shows $r = 23.6$.}
    \label{fig:miniJPAS r}
 \end{figure}

The area of the miniJPAS survey was chosen to overlap with the AEGIS field \citep{Davis_2007_aegis} because in that region there is a wealth of information such as optical spectra from the Baryon Oscillation Spectroscopic Survey (BOSS; \citealt{Dawson_2012}), SDSS and DEEP2/DEEP3 \citep{Cooper_2011, Newman_2013}, as well as X-ray data from XMM. 
However, in our applications we consider only the cross-match of miniJPAS with the SDSS DR12 Superset \citep{SDSSDR12}, which contains visually inspected spectra and redshifts of all BOSS quasar candidates. 
As a result of that cross-match we end up with 117 quasars, 40 galaxies, and 115 stars.
Fig. \ref{fig:miniJPAS r} shows the histograms of the $r$ magnitudes of the objects in the miniJPAS point-like sources sub-sample, as well as the objects from the SDSS cross-match sample, which is also split into the different classes.
The cross-match sample constitutes a ``truth table'' that we can use to check the classification derived on the basis of the miniJPAS J-spectra. 
Although the SDSS cross-match sample constitutes an important test set, one should bear in mind that it is not only extremely small, but it is also biased in terms of brighter sources, stellar types, redshifts, etc. The scarcity of spectroscopically confirmed objects is a problem not only for testing the methods, but mainly for training the ML methods, which require very large data sets in order to tune the weights of the network.
Therefore, in order to train and to validate our classifiers with reliable statistics, we employ simulated data, the mock J-spectra, which are described in the following Section.

\subsection{Mock J-spectra}\label{Data mocks}

ML algorithms are usually trained and validated using real-world data sets, and are subsequently applied to data which are as similar as possible to the training sets.
However, when real data is not available or is too scarce, simulations can be employed to either complement existing real-world training sets, or to build entire training sets -- see, e.g., \citet{2015MNRAS.450..305H}, \citet{syth-z}.

Supervised learning algorithms depend on large and complete training sets with verified labels in order for the models to be properly fitted \citep{5206848}.
In the case of J-PAS/miniJPAS data, the numbers of objects with confirmed labels are barely large enough for us to test the algorithms -- never mind training them. 
Moreover, catalogues of astrophysical objects with confirmed labels are typically biased due to the target selection processes prior to the spectroscopic observations.
They are also typically brighter, allowing for better signal-to-noise ratio (SNR) observations, and as a result may not contain a faithful representation of the variety of objects expected to be found in a deeper, complete sample. 
Therefore, mocks are important in astronomy not only to augment the volume of the training sets, but also to fill in the sample where it lacks in diversity, in terms of magnitude ranges, types and redshifts.

However, the construction of realistic simulated data sets is beset with substantial challenges. 
First, the frequencies of the objects in the training sets need to be kept under control, otherwise we may bias the classes in the validation and test sets. 
Second, the properties of the simulated data itself must mimic, as much as possible, those of the real data sets.
This means that not only the measurements, but their uncertainties, must observe the same distributions in terms of luminosity, object class, and SNR.

In \citet{queiroz2022minijpas} we have described in detail how we have constructed a mock catalogue of quasars, stars and galaxies which reproduce the frequencies of those classes of objects that we expect to find in the real data sets. 
The first step in those simulations is a random sampling of objects drawn from given distribution functions: the quasars obey a standard luminosity function \citep{qsoLF_palanquedelabrouille}, the galaxies follow a distribution based on the miniJPAS sample cross-matched with DEEP3 and SDSS DR12Q used in the quasar selection, and the stellar types and magnitudes follow the distribution expected for the specific region of the Milky Way that overlaps with the AEGIS field \citep{2003A&A...409..523R}. 

After specifying the types, luminosities and redshifts of the objects in the mocks, we search for SDSS optical spectra and compute the fluxes and magnitudes in the J-PAS filters by convolving those spectra with the filters.
The synthetic fluxes are similar to those measured by miniJPAS, except for the fact that their SNRs are typically much higher due to the nature of the SDSS spectroscopic observations.
The next step is, therefore, to add noise to the synthetic fluxes in such a way that the final simulated data set has a SNR distribution which is consistent with the miniJPAS observations.

At this point, care must be taken to reproduce the actual noise properties of the underlying real data set. As shown in \citet{queiroz2022minijpas}, for some filters the noise models turned out not to be well-fitted by a Gaussian, but some are better fitted by slightly different distributions.
In this paper, unless noted otherwise, we train and test our ML methods with the mocks produced using the best-fit noise models, labelled as noise ``model 11''.

Finally, the mocks also model the pattern of non-detections (NDs) from the miniJPAS point-like sources sub-sample. In order to train the ML models, we leave the fluxes exactly as they are in the catalogues, without any special treatment of those low SNR measurements.

We used 4 data sets to train and validate our models.
The training set is a balanced data set \citep{JohnsonK19} containing equal numbers of stars, galaxies and quasars ($10^5$ of each); the validation set, which we used for the ML model selection, contains $10^4$ objects of each class; and the ``balanced test set'' contains another $10^4$ stars, galaxies and quasars. 
In addition, we used an alternative test set, the ``1deg$^2$ test set'', that contains the expected numbers of objects within 1deg$^2$, down to the photometric depths of miniJPAS. Thus, this test set is not balanced.
As usual in ML, both test sets remained completely blind to the training procedure.

\section{Machine Learning Models}\label{ML}

Dividing complex objects into classes is one of the tasks where Machine Learning (ML) has become widely used: identifying letters in written manuscripts, detecting different types of animals in images, or addressing financial risks from socio-economic data, are some of the simplest examples where the applications of ML methods have shown remarkable success.

Here we consider classification using photometric catalogues as the basic data set for classifying the objects, and we focus on the fluxes and their associated errors -- i.e., we will rely on the averaged spectral features of those astrophysical sources.
The set of fluxes (or, equivalently, magnitudes) in broad-band photometric surveys are typically treated as ``tabular data'', since there are only a few measurements that follow a certain order, which can be thought as the central wavelengths of the filters (the photometric bands).

There are in fact some particular ML models that are considered as standard benchmarks for tabular data classification -- e.g. random forests, neural networks, gradient boosting, etc. --, and these methods are also commonly used to separate astrophysical sources \citep{kids_qso_catalog,kids_ML_qso_selection, pbaqui, nakazono}.
In the case of narrow-band surveys, however, we have a significantly higher spectral resolution compared with broad-band surveys. 
This means not only that there is much more data, but that the relevant local features (e.g., breaks, emission and absorption lines) can involve complex combinations of several different points in the input data sequence.

The classification of astrophysical sources involves several additional challenges related to ML such as: biased training sets, handling missing data (e.g. non-observations and non-detections), noisy labels\footnote{By noisy labels we mean objects which have been assigned the wrong class, e.g., galaxies labelled as stars.} and noisy attributes. Moreover, one could also raise the issues of model interpretation and uncertainty quantification.

The problem of biased training sets arises because in astronomy the training sets are usually built based on cross-matches with spectroscopic surveys -- from which we get reliable labels.
Apart from the fact that spectroscopic surveys require significantly more resources compared with imaging, spectroscopic training sets may be biased over bright sources, redshift ranges, etc. 
This is an issue for ML since these models are unreliable on ``out of domain'' samples, i.e., data that extrapolate the training set. 
In this work this problem is partially alleviated with the help of the mocks, which were built not only to increase the size of the training sample, but also to be more representative of what we expect to find in the real data, in terms of brightness, redshift and stellar types.

\citet{queiroz2022minijpas} also avoided the problem of noisy labels as much as possible, by building the mocks only with the sources from the SDSS Superset catalogue, which should return the most reliable classification based on high-resolution spectra complemented by visual inspection.

In this work we also draw special attention to noisy attributes (the errors in input data).
Our catalogues contain, for each object, the 60 fluxes and associated uncertainties provided by the J-PAS filter set.
We test several ML models to classify miniJPAS quasars, stars and galaxies and focus on CNNs because of their flexibility as well as the ease with which we can include the information conveyed by the measurement errors while keeping the context of those errors -- i.e., the fact that a given uncertainty is related to its corresponding measurement \citep{rodrigues2021information}.
These uncertainties inform the significance of individual measurements -- and this is equally true both for template fitting using a $\chi^2$ as for ML methods.
If the data set is very homogeneous, with nearly identical uncertainties for all data points, then of course there is no information in the errors.
But for extremely diverse data sets such as astronomical catalogues, with both bright and faint objects, and a complex distribution of SNRs as a function of magnitude, this information is critical.

We compare the CNNs with two additional ML baseline models: Random Forests (RF, \citealt{breiman2001rf}) and LightGBM (LGBM, \citealt{lgbm}), for which we discard the uncertainties.
Feedback from intrinsically different ML methods gives important hints on how to improve the models, on pre-processing of the input data, and on the validation of the mock data sets. 
In this paper, we also performed a feature importance analysis (Appendix \ref{fi}), which can be used to address the problem of model interpretation.

Regarding pre-processing, in order to pass the inputs to our ML models, we normalise the fluxes and flux uncertainties of any given object according to the root mean square flux for that object:
\begin{align}
\label{eq:norm flux}
\begin{split}
    &f_\lambda \to \frac{f_\lambda}{\sqrt{\sum_\lambda f_\lambda^2}},\\
    &\sigma_\lambda \to \frac{\sigma_\lambda}{\sqrt{\sum_\lambda f_\lambda^2}} \, ,
\end{split}
\end{align}
where the wavelength here is just a label corresponding to the central wavelength of each filter, $\lambda \in (\text{uJAVA, uJPAS, \dots, J1007})$.

In the next subsections we introduce the ML algorithms used in this work.

\subsection{Convolutional Neural Networks}

\begin{figure*}
    \centering
    \includegraphics[width=0.95\linewidth]{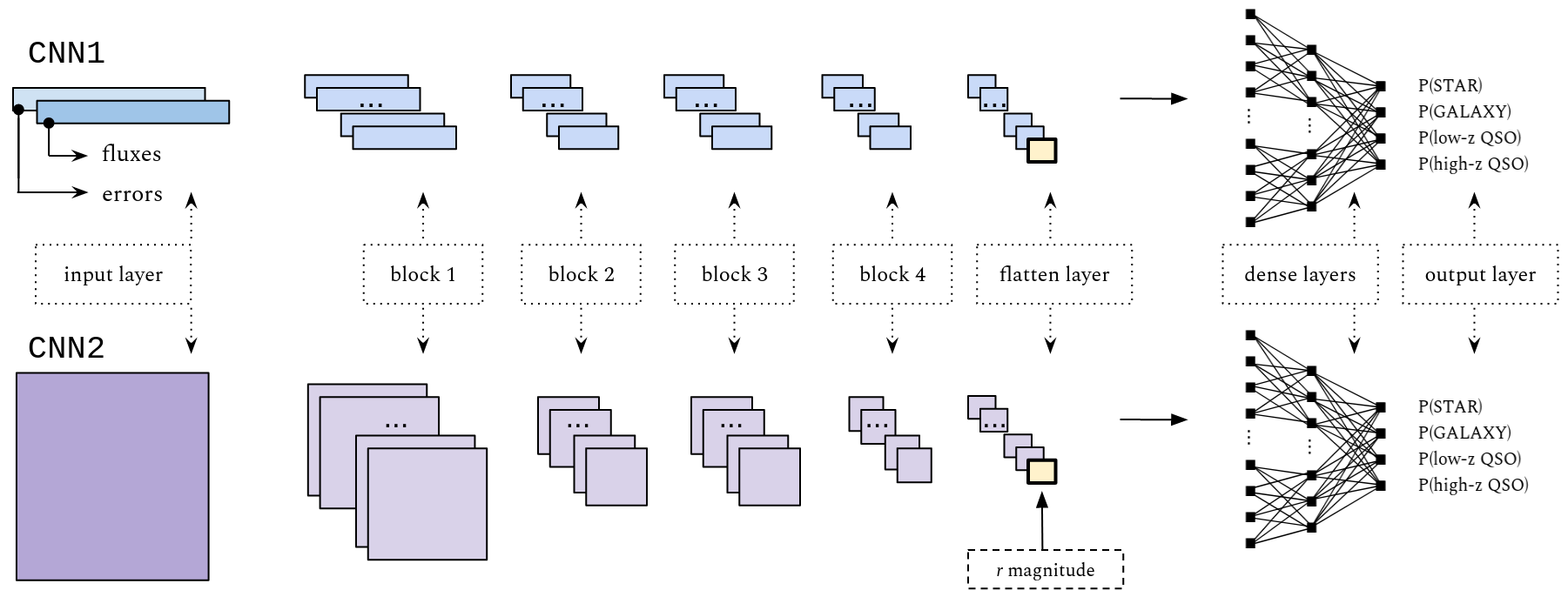}
    \caption{CNN1 (top) and CNN2 (bottom) architectures. 
    CNN1 input data is the set of normalised fluxes and corresponding uncertainties represented as a vector with two channels. 
    One can also train CNN1 without the errors by including only the first channel. 
    The input data of CNN2 is the set of normalised fluxes and corresponding uncertainties represented in 2-dimensions (see Fig. \ref{fig:cnn2 input data}). 
    A ``block'' contains a convolution, batch normalisation, max pooling and dropout layers. 
    The yellow box in the feature maps from both flatten layers represent the $r$ magnitude, which is added to the feature map after the convolution layers have processed the J-spectra.}
    \label{fig:cnn arch}
\end{figure*}

A Neural Network (NN) is a type of learning algorithm where multiple activation units (neurons) are combined through layers to extract information from the data and return a prediction. The input layer receives the set of features of some instance from the data set and to each feature is assigned a weight. 
The activation functions encoded in the neurons from the following layer operate in the scalar product between the features and corresponding weights. 
This procedure is repeated recursively until the last layer, which outputs the predictions.
The layers from NN structures where all neurons are fully connected are called ``dense layers'', and they are designed to learn how to recognise global patterns from the input features.

CNNs work similarly, but were developed to learn how to detect local patterns using convolution kernels. 
For this reason, CNNs have become the benchmark for feature extraction on data sets such as images and sequential data. 
The architectures of CNNs are usually composed of sets of convolution and dense layers: local features are extracted from the input data with the convolution kernels, and are then combined into the dense layers to output the prediction.

In our context, CNNs can be used to search for local features in the J-spectra. 
A similar idea has already been used to classify astrophysical sources from narrow-band surveys in \citet{cabayol2018}, where they show that 1D convolution kernels can be used to classify galaxies and stars, leading to better results when compared to usual ML algorithms, which are due to the ability of the CNNs to extract these local features. 
For an application in the context of high-resolution spectroscopic data see e.g. \citet{busca2018quasarnet, cnn_sfh, cnn_stype}.

We created our own CNN architectures with the help of the \texttt{keras} framework \citep{chollet2015keras}.
We used the \texttt{adam} optimizer to minimize the categorical cross-entropy loss function
\begin{equation}
\label{eq:loss}
    \text{cross-entropy} = -\frac{1}{N}\sum_n^N\sum_k^K y_{nk}\log p_{nk},
\end{equation}
where $N$ is the number of instances, $K$ the number of classes, $y_{nk}$ is the true class and $p_{nk}$ is the assigned probability.
The convergence of the models was monitored using learning curves of the $F_1$ score (see \S\ref{metrics}) and the loss function on the training and validation sets. 
The number of epochs is constrained to the \texttt{EarlyStopping} callback: the training is interrupted when the validation loss stops improving for a number of epochs specified by the \texttt{patience}.
In order to prevent the training from stagnating, we vary the learning rate using the \texttt{ReduceLROnPlateau} callback, which reduces the learning rate when the validation loss stops decreasing for a chosen number of epochs.
We also use the \texttt{ModelCheckPoint} callback to save the set of weights that leads to the best $F_1$ macro-averaged score in the validation set. 
The final model corresponds to this set of weights, ensuring that the model has varied very little in the last epochs.
In all intermediate layers, both convolution and dense, we use as activation the ReLU function $f(x) = \textit{max}(0,\ x)$ \citep{Relubibcode}. 
In the last dense layer, on the other hand, we use the \textit{softmax} activation function in order to obtain a probabilistic interpretation of the output value, i.e., the scores assigned to the four classes add up to one.

The input feature maps and architectures for each CNN version are illustrated in Fig. \ref{fig:cnn arch}. 
We call a set of convolution (\texttt{Conv1D} or \texttt{Conv2D}), \texttt{BatchNormalization}, \texttt{MaxPooling} and \texttt{Dropout} layers a ``block''.

\subsubsection{CNN1}\label{cnn1}

The first CNN version receives as input the set of fluxes (J-spectra) and nominal errors organised as 1D vectors in two channels (upper panel in Fig. \ref{fig:cnn arch}). 
In this way, the learned features from both channels are combined in the output feature map.
We also trained and tested CNN1 without the second channel, i.e., only with the fluxes, without including the uncertainties.

After the set of convolution layers processes the J-spectrum, it returns a tensor which is converted into a one-dimension vector in the \texttt{Flatten} layer. 
In addition to the J-spectrum ``tensor'', we also add as input the $r$ magnitude in the \texttt{Flatten} layer\footnote{With this strategy it is possible to include any other ``tabular'' features, such as morphological parameters or time-domain data, in addition to the J-spectra.}.
This vector then serves as input for two intermediate dense layers with 64 and 32 neurons, which are finally connected to the output layer that returns the scores assigned to each class.

\subsubsection{CNN2}\label{cnn2}
\begin{figure}
    \centering
    \includegraphics[width=1.0\linewidth]{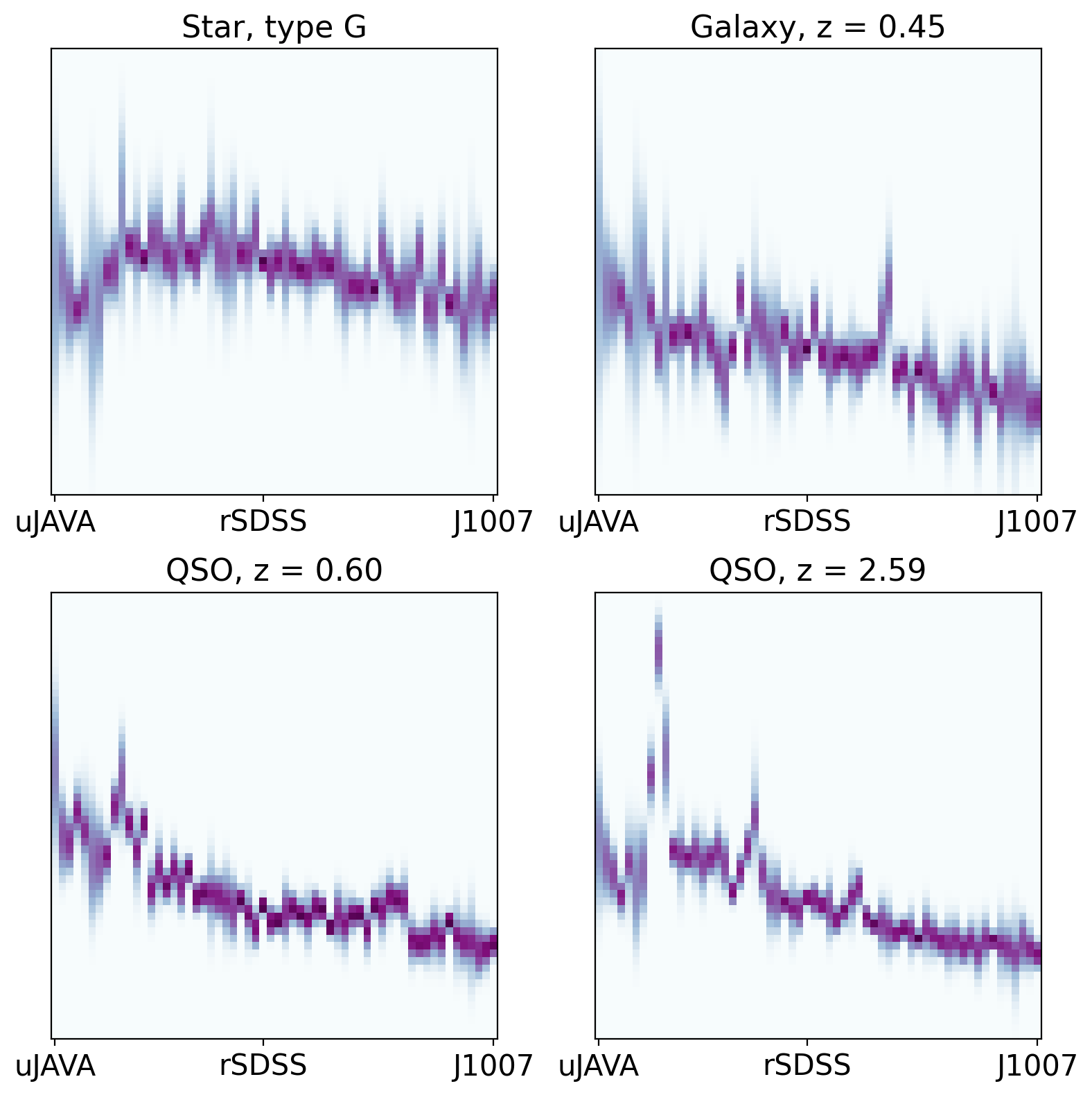}
    \caption{Diagram representing the CNN2 input data. Columns correspond to the miniJPAS filters and rows correspond to normalised fluxes. Darker pixels correspond to higher probability, i.e., denser regions of the probability distribution. The top panels show a G-type star (left) and a galaxy at z = 0.45 (right). The bottom panels show a low-z QSO at z = 0.60 (left) and a high-z QSO at z = 2.59 (right). Computed according to noise model 11.}
    \label{fig:cnn2 input data}
\end{figure}

The strategy used to account for the uncertainties as input features in CNN2 is to treat the measurements as probability distributions with mean value equal to the flux measurements, and standard deviation equal to the corresponding nominal errors  \citep{rodrigues2021information}. 
These distributions are then represented as 2-dimensional matrices, as illustrated in Fig. \ref{fig:cnn2 input data}.
This format for the input data can be particularly useful to represent errors which do not follow a simple form such as a Gaussian distribution.
Furthemore, since the matrix representation is identical to an image, it is naturally suited for CNNs with 2D convolution kernels.
The idea of representing fluxes and uncertainties as heatmaps have already been used in the context of astrophysical sources classification \citep{scone, helenqu_22}.
The bottom panel of Fig. \ref{fig:cnn arch} illustrates the architecture of our CNN2 method. 
Once again, we add the $r$ magnitude in the \texttt{Flatten} layer feature map, and the dense layers contain 64 and 32 neurons, as in CNN1.

\subsection{Decision Tree Based Algorithms}\label{ML DT}

In the following subsections we introduce the decision tree based models used to compare with the performance of the CNNs. 
The details about hyperparameter (HP) tuning of these models are described in Appendix \ref{hp tune}.

A DT is a structure where the algorithm makes predictions by splitting the data set based on constraints imposed in terms of the features. 
Each decision rule is encoded in a node of the tree. 
The algorithm establishes which feature will be evaluated at each node by measuring the worth of a split based on each of the features. 
This is quantified by the information gain, which measures the expected decrease in some impurity function. 
This function can be either \textit{entropy} or the \textit{gini impurity}. 
The features which lead to the highest increase in the gain are then allocated to the corresponding node.

\subsubsection{Random Forests}

Random Forests (RF) have been widely used for many tasks related to astrophysical data, including source classification \citep{kids_qso_catalog, pbaqui, nakazono}.
The method consists of combining multiple decision trees to avoid overfitting and build a powerful classifier. 

We implemented the RF model with the \texttt{scikit-learn} \citep{scikit-learn} python package.
Each tree is built with a sub-sample of the data, using the bootstrap aggregating (\textit{bagging}, \citealt{breiman96_bagging}) technique.

The number of features to consider when looking for the best split is by default set as the square root of the total number of features. 
The mechanism of combining independent trees using the bagging strategy makes RF robust to overfitting, and is usually not necessary to limit the growth of each individual tree.

The size of the sub-sample, the number of features and the maximum depth of the trees are examples of RF HPs. 
The chosen values of the HPs from \texttt{scikit-learn RandomForestClassifier} are specified in Table \ref{tab:rf hps}.

\subsubsection{LightGBM}
Gradient Boosting Decision Tree (GBDT) is another type of DT ensemble method which has also proved to be an excellent tool for a variety of problems, including astrophysical source classification  \citep{kids_qso_catalog}.
As opposed to RF, the trees are not grown independently. 
Instead, each tree is built to reduce the error of the previous one. 
This is an iterative method that uses gradient descent to minimize the loss function, which we chose to be the categorical cross-entropy -- see Eq.\eqref{eq:loss}.

We implemented GBDTs with LightGBM (LGBM, \citealt{lgbm}).
There are several well-succeeded frameworks to implement GBDTs, for example XGBoost \citep{xgboost}. 
LGBM was developed to accelerate the training, but it often presents similar (or even better) performance compared with XGBoost.
However, due to LGBM's leaf-wise growth scheme, it might be susceptible to overfitting, so we limit the growth rate and the maximum number of leaves of the trees (see Table \ref{tab:lgbm hps}).

\section{Performance in the mock test sets}\label{results mocks}

We start analysing the performance of the CNN1 (with and without the errors), CNN2, RF and LGBM classifiers when they are applied to the mock test samples. 
The results when applying those methods to real data will be shown in the next Section.

\subsection{Evaluation Metrics}\label{metrics}

In order to build a high quality quasar catalogue we need to find the best possible balance between \textit{completeness} and \textit{purity}, i.e., we want to recover the highest fraction of quasars possible, but in a such a way that our sample remains as free from contaminants as  possible. 
With that in mind, we evaluate the performance of the classifiers by computing both purity (``precision'') and completeness (``recall''):
\begin{align}
    \textit{purity} &= \frac{\text{TP}}{\text{TP} + \text{FP}} \, , \\
    \textit{completeness} &= \frac{\text{TP}}{\text{TP} + \text{FN}} \, ,
\end{align}
where TP, FP and FN are true positive, false positive and false negative, respectively.
In order to find the ideal balance between completeness and purity, it is useful to define the $F_1$ score, which combines both scores into a single number:
\begin{equation}
    F_1 = 2 \times \frac{\textit{purity} \times \textit{completeness}}{\textit{purity} + \textit{completeness}} \, .
\end{equation}

All ML models employed in this work return a score associated with each class, which can be interpreted as a proxy for the probability that an object belongs to that class. The scores of all classes add up to 1.
We have the freedom to choose different thresholds for these classification scores (the ``probabilities'') in order to improve the final classification. 
Depending on that choice, one may obtain a more complete or more pure sample, i.e., the $F_1$ score depends on the threshold. 
By default, the chosen class $k$ corresponds to the class with the highest score, according to the argmax rule:
\begin{equation}
\label{eq:argmax}
    y_i = \underset{k}{\text{argmax}} \, f^{k}(\mathbf{x}_i) \, ,
\end{equation}
where $\mathbf{x}_i, y_i$ are the input data and predicted class of instance $i$, respectively, and $f$ is some function that assigns probabilities to each class $k$. This means that, when we apply some trained ML model to classify an instance $i$, it returns a probability associated to each class $k$ and the final class correspond to $k$ with highest score.

Another useful metric is the \textit{Receiver Operating Characteristic} (ROC) curve, because it shows the quality of a classifier before choosing a specific threshold by computing the true positive rate \textit{versus} false positive rate. 
Moreover, the \textit{Area Under the ROC Curve} (ROC-AUC) is a useful summary statistic of the ROC curve to measure the quality of a classifier. 
Since we are working with multiple classes, we computed the one-\textit{versus}-all ROC-AUC score.

Finally, in order to compare the overall performance of a classifier by considering the performance over all classes, it is useful to compute the unweighted, or \textit{macro} averaged score, defined as:
\begin{equation}
\label{eq:f1 macro}
    \Bar{S} = \frac{1}{K}\sum_{k}^{K}S_k \, ,
\end{equation}
where $S$ is some score or metric, $k$ labels the individual classes and $K$ is the total number of classes. 
This metric does not take into account the imbalance of classes, and thus avoids biasing the analysis over more frequent types.

\begin{figure}
    \centering
    \includegraphics[width=1.0\linewidth]{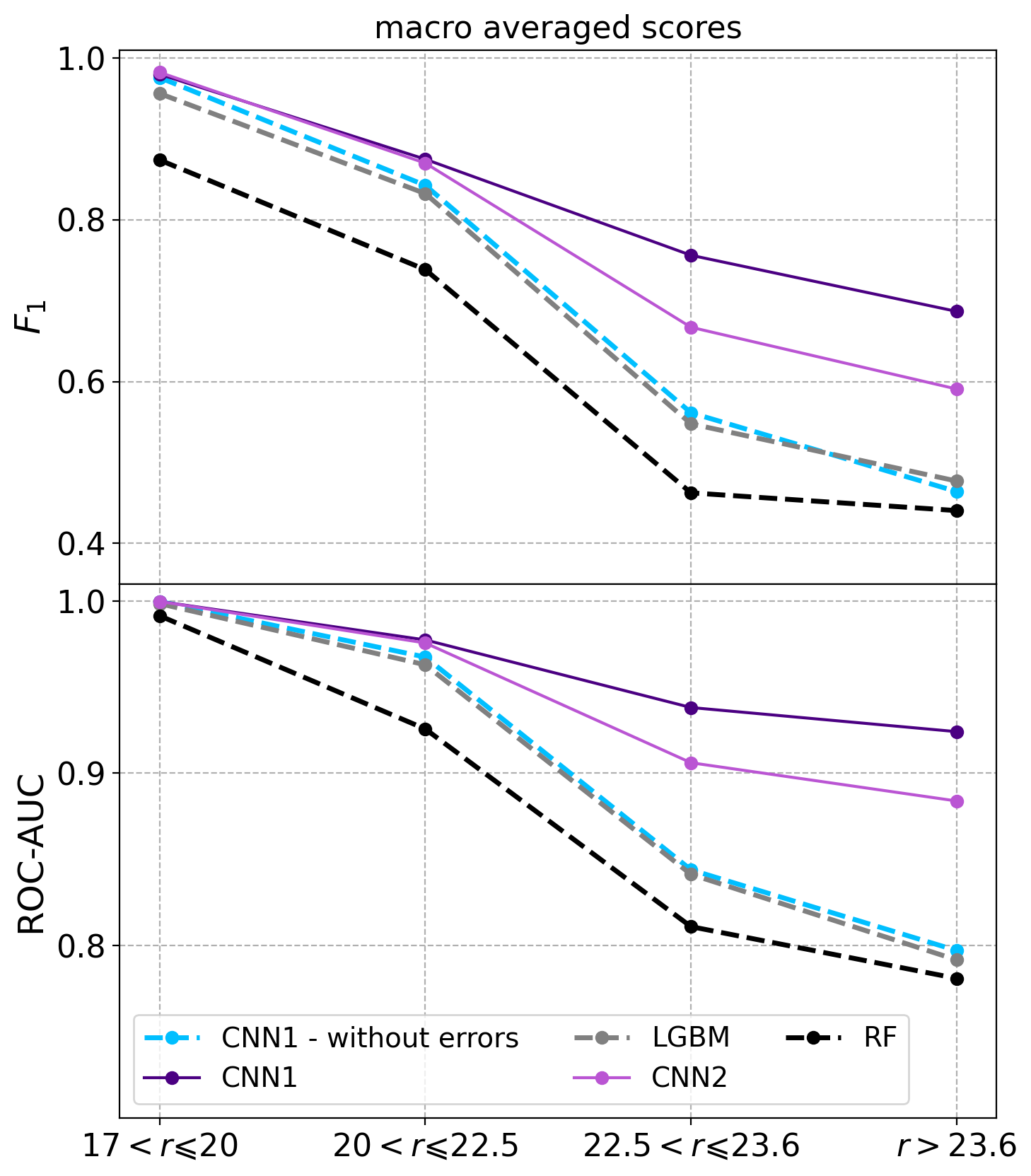}
    \caption{Performance of the ML models when applied to the balanced test set, in terms of the macro averaged $F_1$ score (top), and the ROC-AUC (bottom), for the different $r$-magnitude bins.}
    \label{fig:compare ML}
\end{figure}

Fig. \ref{fig:compare ML} shows the macro-averaged $F_1$ and ROC-AUC scores obtained with the classifiers in the balanced test set, in multiple intervals of $r$ magnitude -- see Appendix \ref{app results} for the complete confusion matrices. 
It is striking how much the performance of the CNN1 classifier improves when the information about the errors is included, in particular for the fainter objects where SNR is even more crucial.
That performance is similar using CNN2, which employs an entirely different architecture for the input data but that, like CNN1, also uses the convolutional layers to incorporate the errors in the context of their corresponding measurements.
The fact that both DT-based methods (specially LGBM), which do not take the errors into account, attain a performance that is similar to CNN1 without errors indicates that the reason for the improvement in the classification seen in the two CNN methods with errors is in fact due to the additional information contained in the uncertainties.
We also see from Fig. \ref{fig:compare ML} that the performances of all the classifiers degrade as the samples become fainter, which is expected since those objects are increasingly noisier and therefore harder to identify. We used the same magnitude bins as Martínez-Solaeche et al. (in preparation), which verified a similar behaviour.
Due to its superior performance, we will focus on the results obtained with CNN1 for the remainder of this Section, unless noted otherwise -- but we emphasise that the results outlined here are qualitatively consistent between all classifiers.

\subsection{Results}\label{results mocks subsection}

\begin{figure*}
    \centering
    \includegraphics[width=1.0\linewidth]{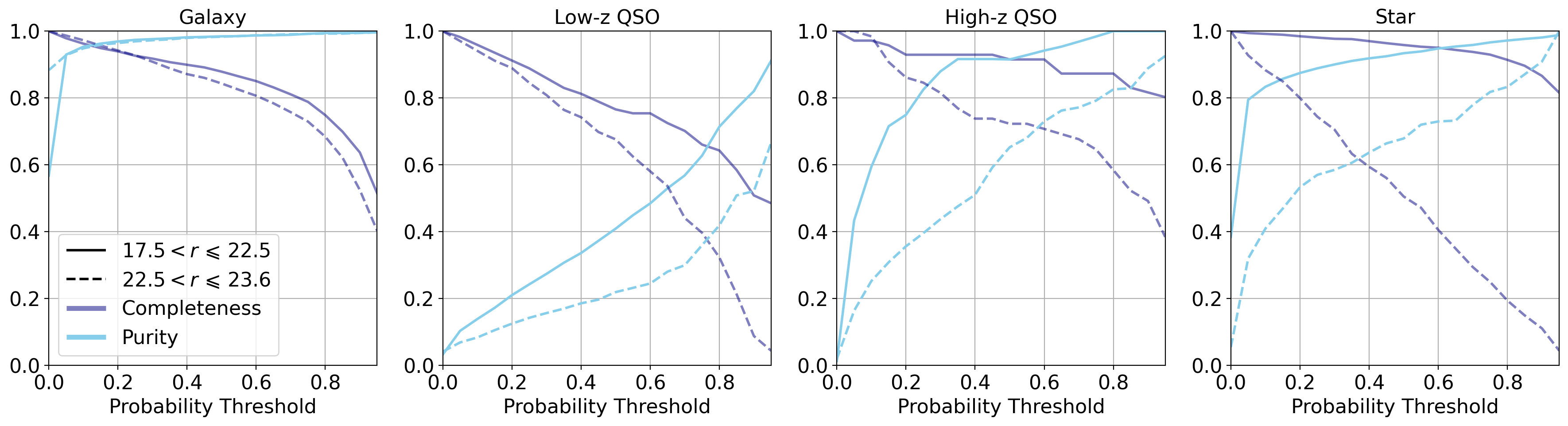}
    \caption{Completeness and purity of the CNN1 method for the mock 1deg$^2$ test, as a function of the probability threshold, for each class. Brighter (fainter) objects are shown in solid (dashed) lines.
    }
    \label{fig:prob cuts}
\end{figure*}

In this section we present the results when we apply the CNN1 method to the two mock test sets: the balanced test set (with $10^4$ objects in each class), and the 1 deg$^2$ test set, which is perhaps a more realistic representation of the miniJPAS sample. 
For much of this analysis, it is more revealing to evaluate the predictions in terms of the balanced test set, just because it is the largest one and we can thus work with more reliable statistics.
However, evaluating the proper choice of threshold using the balanced test set can be misleading, since we want to estimate the purity and completeness in a realistic scenario, with the expected fraction of objects of each class.
Therefore, we start by showing, in Fig. \ref{fig:prob cuts}, the purity and completeness as a function of the probability threshold in the 1deg$^2$ test set. 
We split the sample in two bins of $r$ magnitude, $17.5 < r \leq 22.5$ and $22.5 < r \leq 23.6$, because the optimal choice for the cut might depend on how bright the object is: fainter objects are much noisier, so we expect a classifier to be less confident in this regime.
Based on this analysis, we define the ``1deg$^2$ threshold criteria'' to select candidates in the miniJPAS catalog -- one value for bright and one for faint sources, according to the magnitude bins showed in Fig. \ref{fig:prob cuts}. It corresponds to the value of threshold that leads to highest $F_1$ score in the 1deg$^2$ test sample, and it must be at least equal to 0.5 to ensure that the probability associated to some class is greater than the sum of the others. 
Notice, however, that not all objects are assigned a class according to this criterium.

For the remaining of the analysis in this Section, we work with the balanced test set, for which the best choice of threshold is in good approximation the ``argmax'' criterium, defined in Eq.\eqref{eq:argmax}.

\begin{figure}
    \centering
    \includegraphics[width=0.85\linewidth]{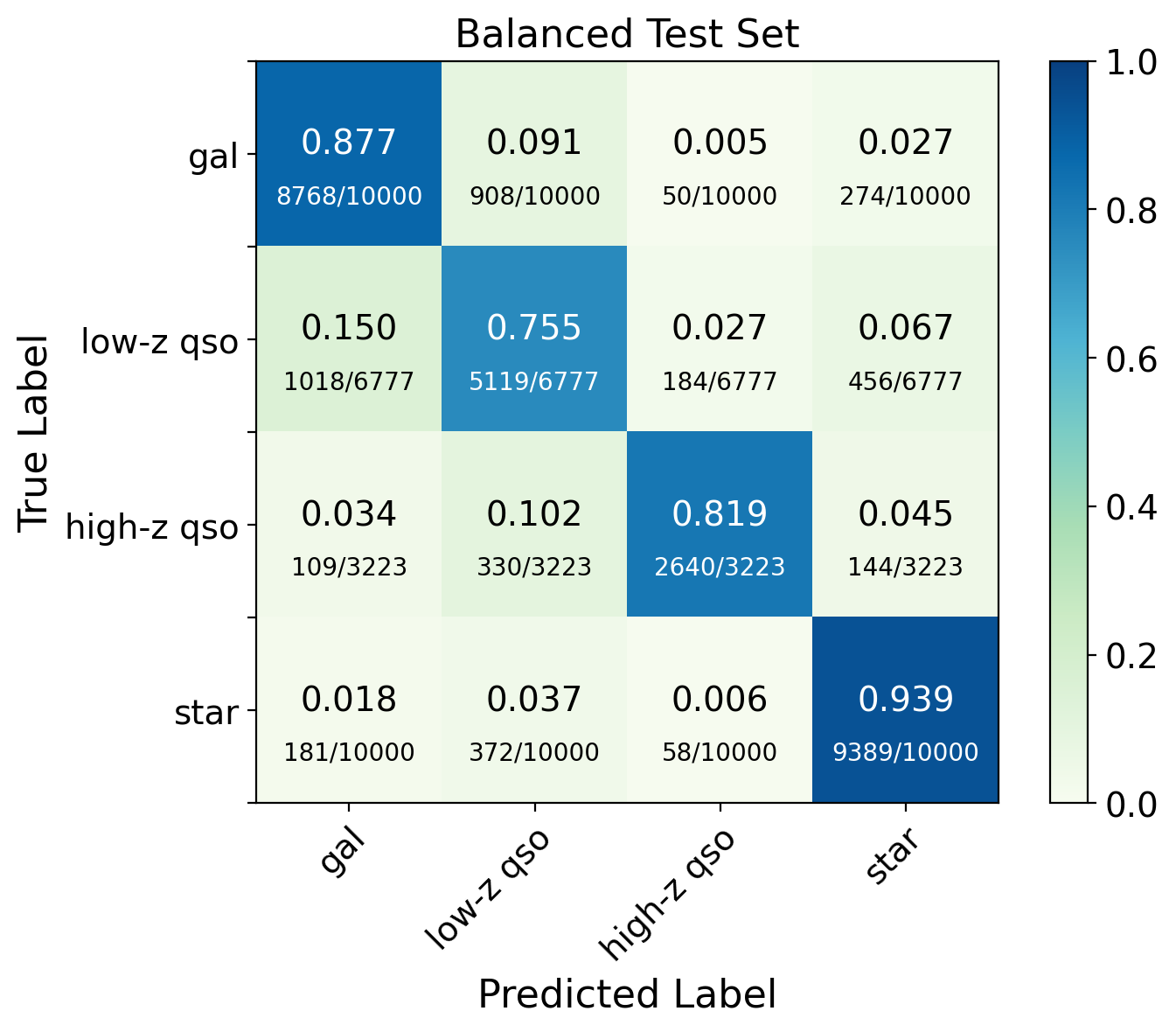}
    \caption{Confusion matrix computed with CNN1 for the mock balanced test.}
    \label{fig:overall CM CNN2}
\end{figure}

Fig. \ref{fig:overall CM CNN2} shows the confusion matrix computed with CNN1. 
We are able to distinguish between low-z and high-z QSOs satisfactorily, and the main source of confusion is between low-z QSOs and galaxies, which is in agreement with the results of Martínez-Solaeche et al. (in preparation).
In Appendix \ref{app results} we show the confusion matrices split in the same $r$-magnitude bins as in Fig. \ref{fig:compare ML}, for all ML methods.

As a complementary analysis, we trained CNN1 in a binary classification scheme, by labelling low-z and high-z QSOs as one single class, and stars together with galaxies as another class. 
The results of that analysis are nearly identical with the numbers shown in Fig. \ref{fig:overall CM CNN2} when we combine the low-z and high-z QSOs in one class, and the stars and galaxies in the other class.

\begin{figure}
    \centering
    \includegraphics[width=1.0\linewidth]{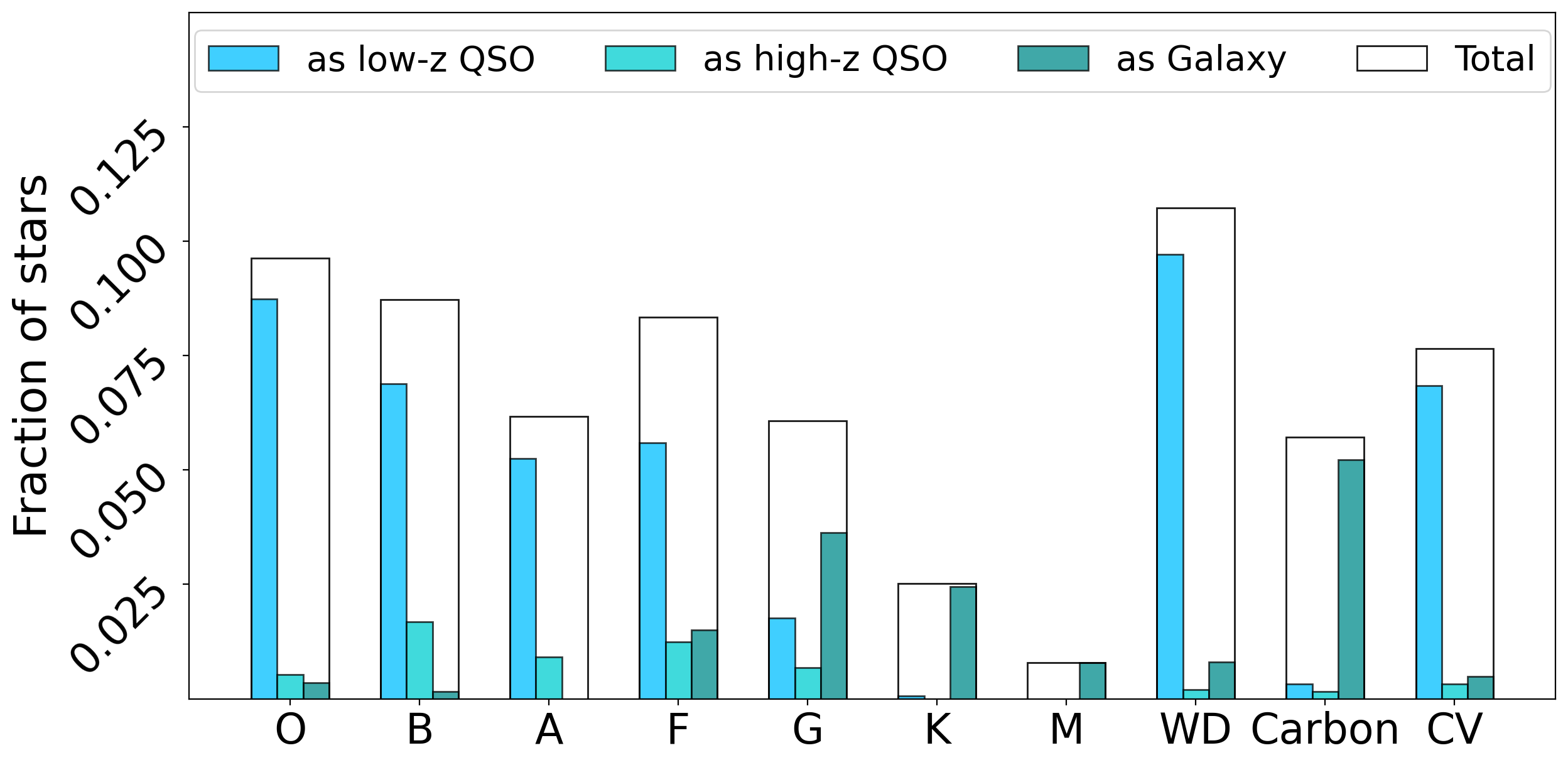}
    \caption{Fraction of stellar types that were incorrectly classified by CNN1 in the balanced test set.
    \label{fig:wrong stars}}
\end{figure}

\begin{figure}
    \centering
    \includegraphics[width=0.8\linewidth]{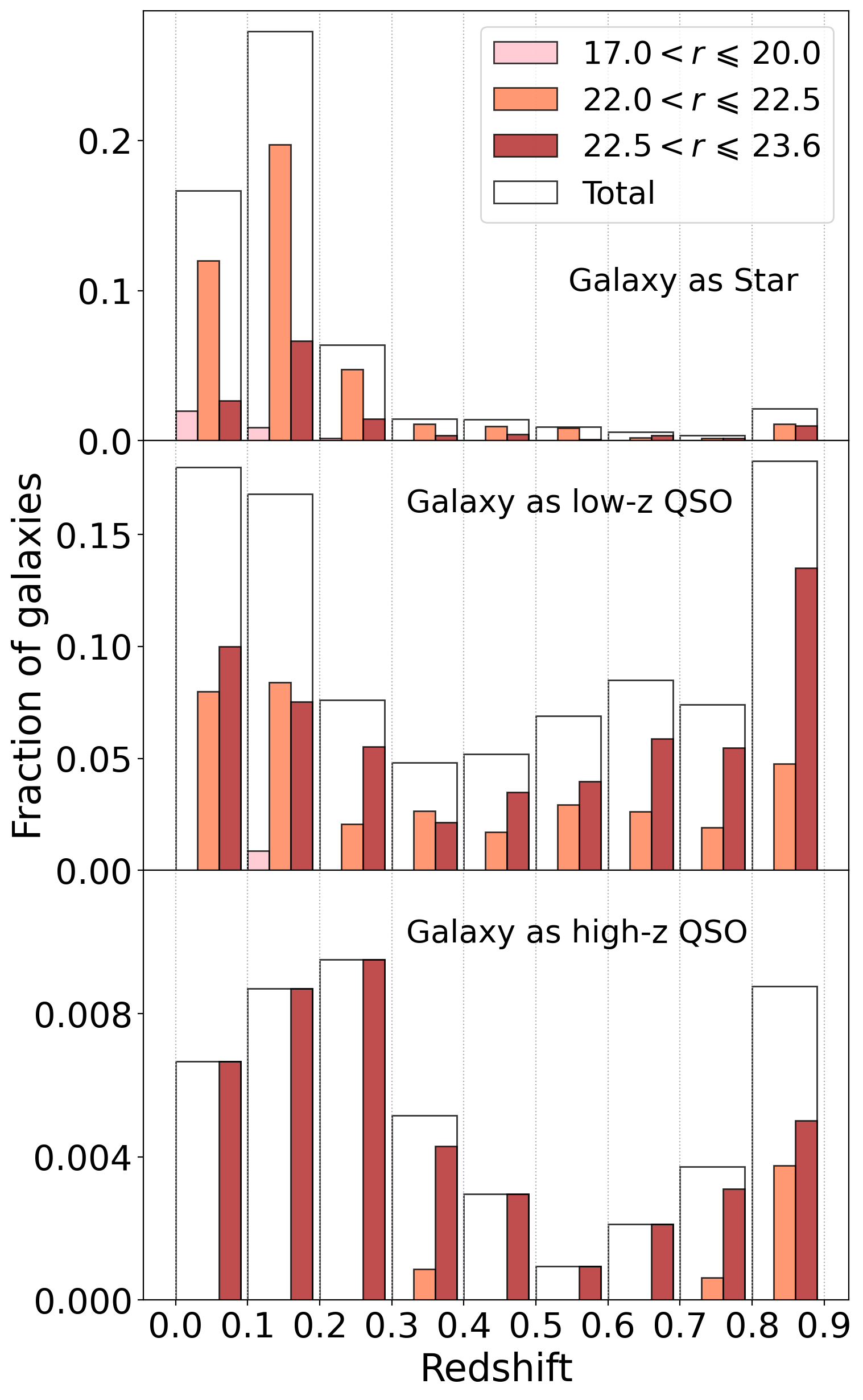}
    \caption{Redshift of the galaxies that were incorrectly classified as stars (top), low-z QSOs (middle) and high-z QSOs (bottom) by CNN1 in the balanced test set.
    \label{fig:wrong gals}}
\end{figure}

\begin{figure}
    \centering
    \includegraphics[width=1.0\linewidth]{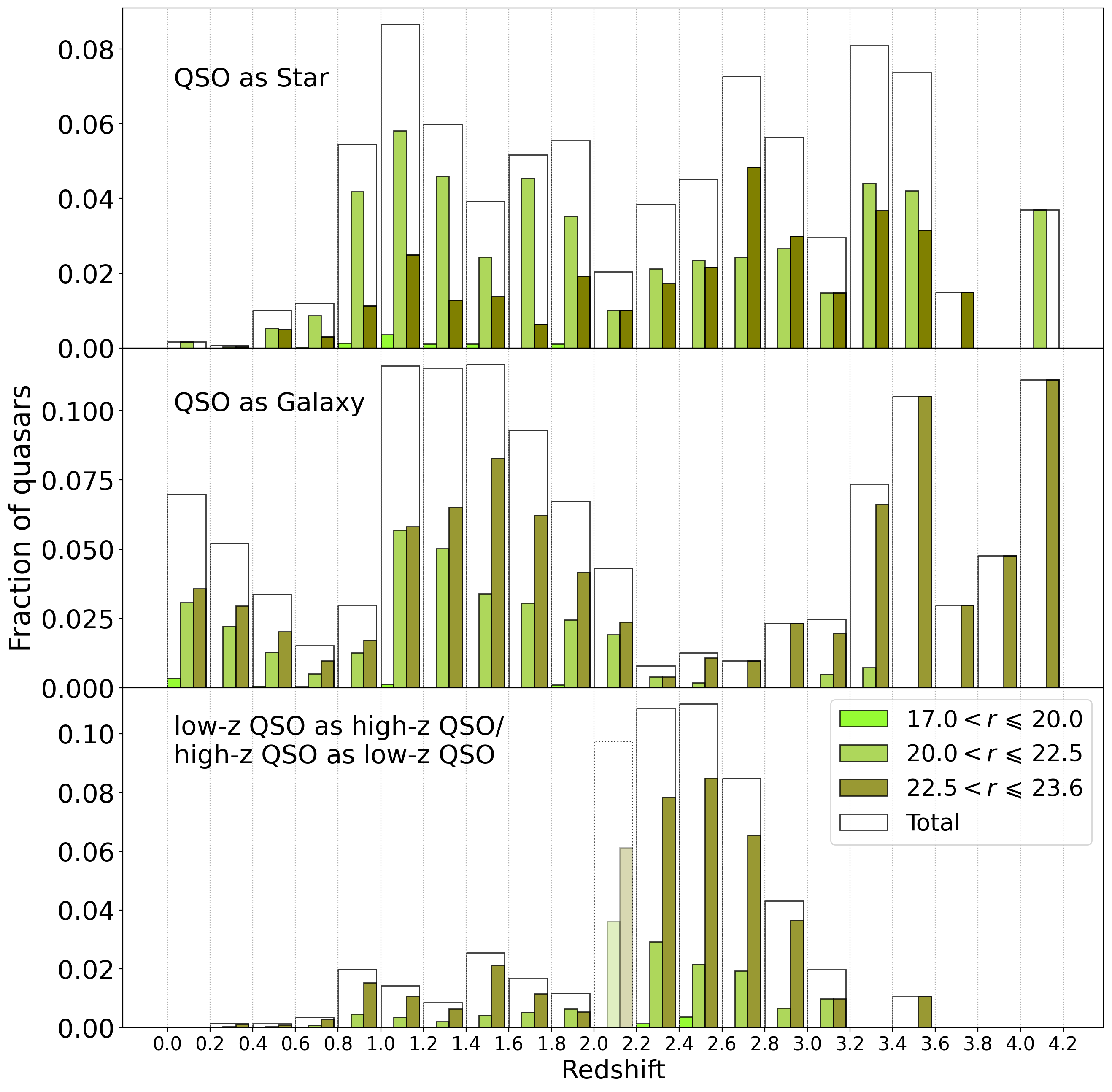}
    \caption{Redshifts of the quasars from the balanced test set that were incorrectly classified by CNN1. The top and middle panels show the quasars which were classified as stars and galaxies, respectively, and the bottom panel shows the quasars which were classified as quasars in the wrong redshift interval. The bars of the histograms cover a redshift range of $\Delta z = 0.2$ and were split according to the $r$ magnitude. The bin containing the pivot value $z = 2.1$ that separates low-z QSO from high-z QSO is shown with different transparency.}
    \label{fig:wrong qso redshift}
\end{figure}

Fig. \ref{fig:wrong stars} shows the fractions of stellar types that were incorrectly classified in the mock balanced test set. 
We show this result in terms of fractions to avoid biasing the analysis over more frequent stellar types, i.e., we take the ratio between the number of incorrectly classified stars of a given stellar type and the total number of stars of the corresponding type.
White dwarfs (WD) and O type stars show the highest fraction of incorrect classifications, which are often classified as low-z QSOs.
The steep blue continuum of the WD spectra can be easily mistaken for the blue and featureless continuum of the QSO spectra at low redshifts \citep{Richards_2002,Myers2015ApJS}.
The A, F and M stellar types also have colours and/or spectral features similar to those from QSOs. However we did not detect significant confusion of these types as compared to the others. In particular, M stars are classified as galaxies. 
Stars of types O, B and F are featureless, which might explain the significant confusion with low-z QSOs. 
In fact, low-z QSOs is the class of objects that is most affected by contamination from stars, which is a well-known problem for broad-band classification in the optical range \citep{2009ApJS..180...67R}, that still persists even with narrow-band data.
For the few stars that end up classified as high-z QSOs, most are of types B, A, F as well as some G stars, although some stellar types outside the main sequence (WD, Carbon and CV) can also contaminate that sample. 
In general, redder stars (G, K, M) are more often confused with galaxies, while bluer stars (O, B, A, F) are more often confused with QSOs, and Carbon stars is the type most often confused with galaxies.

In Fig. \ref{fig:wrong gals} we show the redshifts of the galaxies for each magnitude bin that are confused with stars (upper), low-z QSOs (middle) and high-z QSOs (lower panel).
Since galaxies and quasars are typically not as bright as Milky Way stars, we split the samples into bins of magnitude in the $r$ band in order to check the dependence of the classification on the brightness of these sources. Once again, we compute the fraction of galaxies, now in each redshift bin.
There is very little leaking of galaxies to high-z quasars and it is dominated by fainter objects. From the confusion matrix in Fig.\ref{fig:overall CM CNN2} we see there are only 50 galaxies classified as high-z QSOs.
The galaxies which are classified as low-z QSOs (and also those classified as stars) have typically lower redshifts, although we see similarly high confusion of galaxies within $0.8 < z < 0.9$ and low-z QSOs.

In Fig. \ref{fig:wrong qso redshift} we show the redshifts of the QSOs that were incorrectly classified. Similarly to what happened for the incorrectly classified galaxies, the confusion as a function of redshift is partially related to the fainter magnitudes of these objects. 
The top and middle panels show the QSOs which were classified as stars and galaxies, respectively. The bottom panel shows the low-z/high-z QSOs which were classified as high-z/low-z QSOs.

QSOs classified as galaxies are typically fainter, while those classified as stars are similarly distributed in the bright and faint ends. This reflects the fact that, on the one hand, faint objects are harder to classify, and we thus expect a higher mixing at this regime. On the other hand, stars are most abundant in the bright end, and therefore are expected to be the most frequent contaminants.

The quasar population within $z \in [0.6, 2.0]$ has a scarcity of emission lines, which could explain the confusion with stars and galaxies. 
For $z < 0.6$ we see a higher contamination of the galaxy sample that does not happen for stars. 
This might be due to the fact that the strongest QSO emission lines in this redshift range, such as H$\alpha$, are also commonly found in galaxies.

From Fig.\ref{fig:overall CM CNN2} we see that 10\% of the high-z QSOs are classified as low-z QSOs, while only 2.7\% of low-z QSOs are classified as high-z QSOs. The bottom panel of Fig. \ref{fig:wrong qso redshift} shows the redshift of those objects.
The redshift cut at $z = 2.1$, which differentiates the two populations of QSOs, blurs the distinction between the two classes in the $z \in [2.0, 2.2]$ range (the bin with higher transparency).
The number of incorrectly high-z QSOs that are classified as low-z QSOs starts dropping for $z>2.2$, faster for bright objects and slower for faint objects, indicating some level of confusion between the Ly-$\alpha$ break and other spectral features of the low-z quasars.

\subsection{Robustness tests}\label{robust tests}

\begin{figure*}
    \centering
    \includegraphics[width=1.0\linewidth]{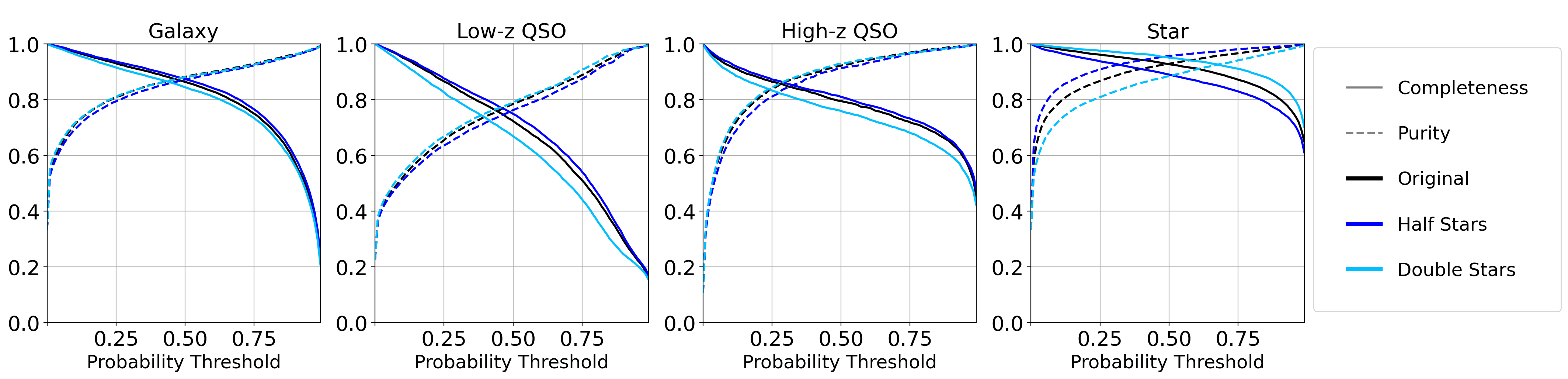}
    \caption{Robustness tests performed with CNN1 in the balanced test set. Completeness (solid lines) and purity (dashed lines) as a function of the probability threshold for each class. Different colours represent different training sets.}
    \label{fig:vary train set}
\end{figure*}

We tested the robustness of the ML classification by modifying the composition of the training set in different ways. 
After training the ML model with the modified samples, we evaluated the performance in the balanced test set, which we kept unchanged.

The first test consists in using only 50\% of the original numbers of stars and keeping the number of galaxies and QSOs from the original training set (\textit{half stars} test). 
Since the stars were removed randomly, we expect the completeness of the star sample and, as consequence, the purity of the other classes, to decrease to some extent. 
The second exercise is the \textit{double stars} test, which is complementary to the previous one: we exclude 50\% of the galaxies and 50\% of the quasars (once again, randomly), while maintaining all the stars of the original training set. 

The idea is that, by changing the proportion of classes in the training set, we expect the models to show a drop in their performances, in particular for the less-represented types. 
If the classification is very sensitive to small changes in the exact mixes of populations in the training set, then the model is not robust.
If, on the other hand, the performance of the classifier drops by only a small amount after a significant change to the training set, then the ML model has converged to a nearly stationary regime.

Fig. \ref{fig:vary train set} shows the scores (completeness and purity) as a function of the probability threshold for the different training sets. 
As expected, the completeness of stars drops in the \textit{half stars} test. 
Although the purity of stars increases, it does not compensate the loss in the completeness, which also translates into a lower purity of galaxies and quasars (especially at high redshift). 
The same reasoning works for the \textit{double stars} test: the completeness of quasars and galaxies drops by a small amount, but there is no significant gain in the purity because of the mixing between these two classes.

These results reflect a well-known feature of ML techniques, which are unable to reliably identify objects that are poorly represented in the training set.
Nevertheless, we verified that the performance of our classifiers is relatively insensitive to significant changes in the training sample, which indicates our ML models are robust in that sense.

\section{miniJPAS point-like sources classification}\label{results miniJPAS}

\begin{figure}
    \centering
    \includegraphics[width=0.85\linewidth]{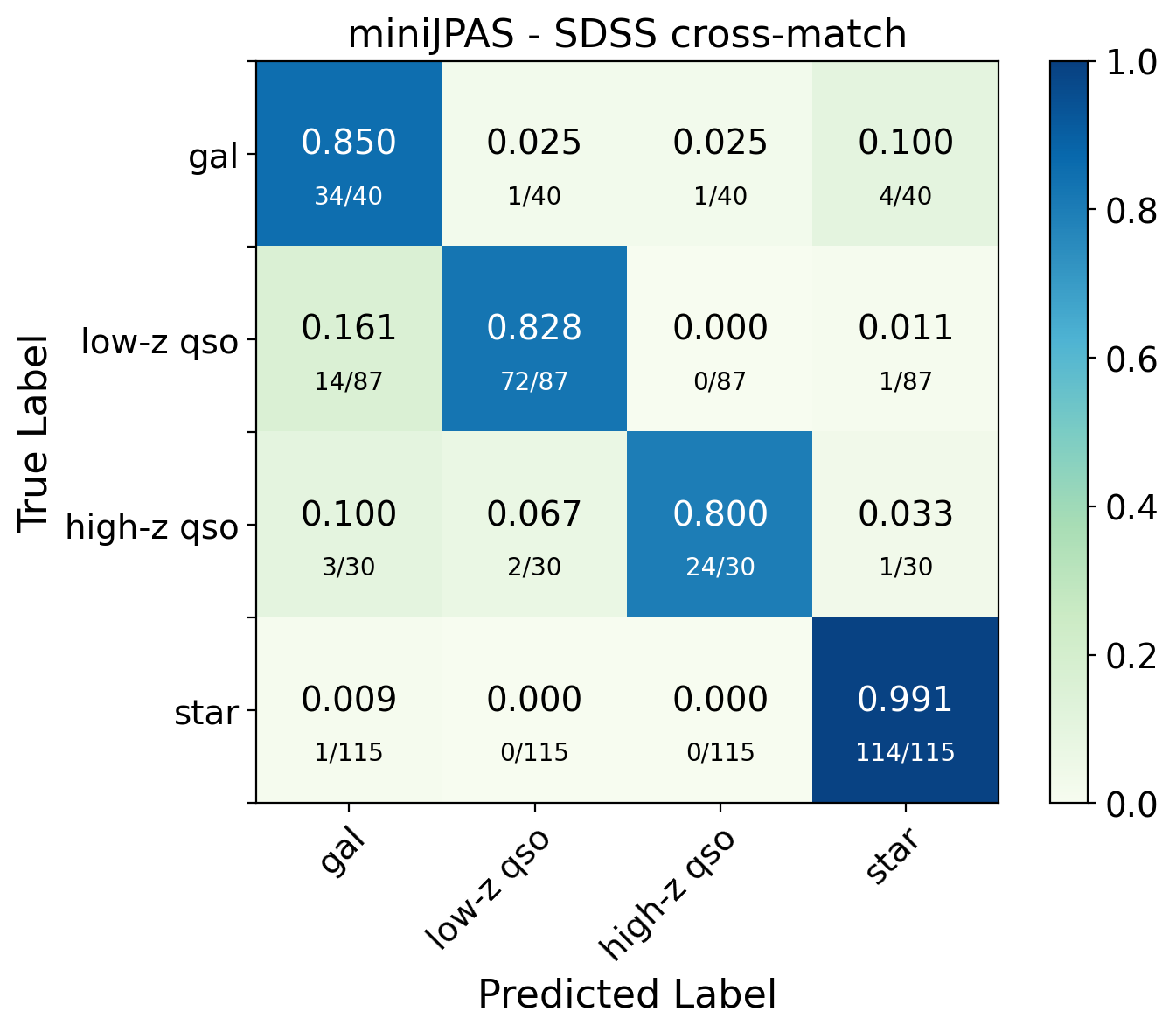}
    \caption{Confusion matrix obtained with the method CNN1 for the cross-match of the miniJPAS point sources with the SDSS DR12 Superset.}
    \label{fig:sdss CM CNN2}
 \end{figure}

In this Section we discuss the predictions in the miniJPAS point-like sources subsample, which contains 11,419 objects. Considering only the magnitude range of $17.5 \leq r < 23.6$, we are left with 7,468 sources.
We have spectroscopic confirmation for some of the objects in this data set, obtained by cross-matching the miniJPAS catalog with the SDSS DR12 Superset (see \S\ref{Data miniJPAS}).
The confusion matrix obtained for that sample is shown in Fig. \ref{fig:sdss CM CNN2}.
The typical magnitude range covered by this sample is $17.5 \leq r \leq 22.5$ (see Fig.\ref{fig:miniJPAS r}). Therefore, in order to see the degradation of the results on real data relative to the mocks, one should compare Fig.\ref{fig:sdss CM CNN2} with the first two bins from Fig.\ref{fig:cm rbins}.
The completeness of all classes is higher than 0.8, which is a good indication that the models trained with the mocks translate fairly well to real data predictions. 
In particular, we see, once again, that the main source of confusion is between low-z QSOs and galaxies.
Regarding high-z QSOs, of the 30 objects of the cross-match sample, 3 were incorrectly classified as galaxies, 2 as low-z QSOs, and only 1 as a star.

\begin{figure*}
    \centering
    \includegraphics[width=1.0\linewidth]{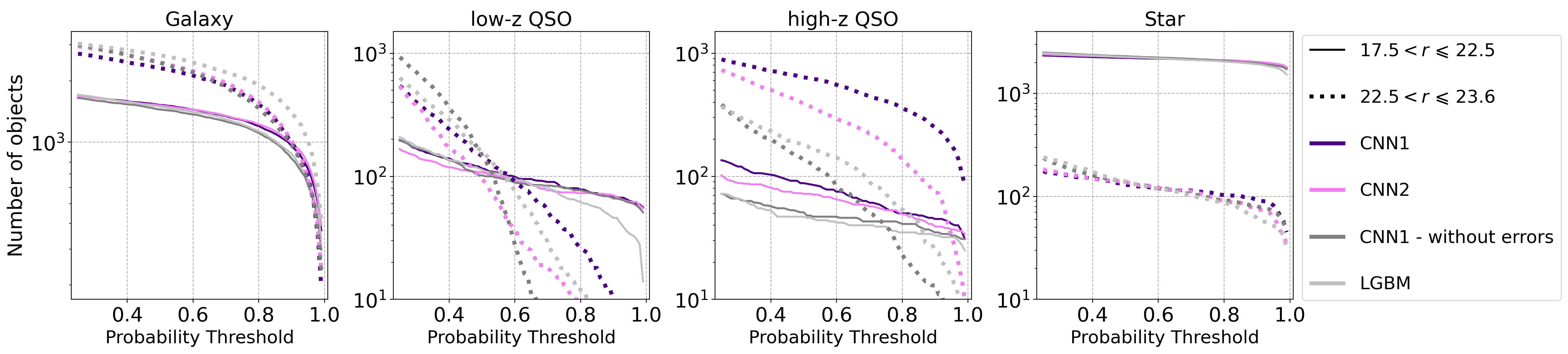}
    \caption{Number of objects predicted by different classifiers as a function of the probability threshold. We compare models which do (coloured lines) and do not (gray lines) include the uncertainties: CNN1 (purple), CNN2 (pink), CNN1 without errors (dark gray) and LGBM (light gray). Solid and dotted lines represent objects in different $r$ bins.}
    \label{fig:miniJPAS model1 and model11}
 \end{figure*}

Fig. \ref{fig:miniJPAS model1 and model11} shows the number of objects found in the point-like sources catalog within $r \in [17.5, 23.6]$, as a function of the ML score (``probability'') threshold.
Coloured lines show the models which include the uncertainties (CNN1 and CNN2) and, for comparison, the gray lines show LGBM and CNN1 without errors.
The choice of threshold for CNN1 can be guided by Fig. \ref{fig:prob cuts}.
Once again, we split the sample into brighter and fainter objects ($17.5< r \leq 22.5$ and $22.5< r \leq 23.6$, respectively) in order to evaluate how many sources of each class are found in these two regimes, and to evaluate how confident the models are when facing brighter and fainter objects. 
The dotted curves show a more dramatic decrease for higher probabilities, which means that the models are less confident when presented with fainter objects.
According to the ML classifiers, stars (galaxies) are the most abundant objects in the bright (faint) end.
The numbers predicted by the classifiers are very similar for bright objects.
CNN1 and CNN2, however, find a significantly higher number of faint high-z QSO as compared to LGBM and CNN1 without errors.

\begin{figure*}
    \centering
    \includegraphics[width=1.0\linewidth]{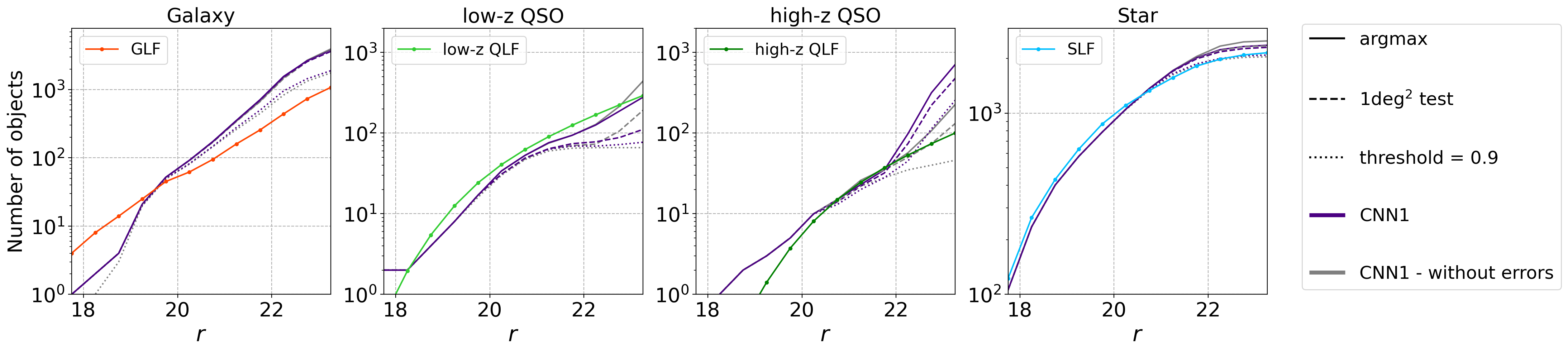}
    \caption{Number of objects predicted by CNN1 with (purple) and without (gray) errors as a function of the $r$ magnitude. We compare the obtained numbers when imposing different probability threshold criteria: argmax (solid lines), 1deg$^2$ (dashed lines) and a very strict choice of threshold = 0.9 (dotted lines). The luminosity functions (LF) of each type are shown as coloured solid-dotted lines for comparison.
    }
    \label{fig:miniJPAS lum function}
 \end{figure*}

Fig. \ref{fig:miniJPAS lum function} shows the number of objects classified by CNN1 with and without errors, along with the number predicted by the corresponding luminosity functions (LF, see \S2.2 and also \citealt{queiroz2022minijpas}).
We show both the classification obtained using the argmax rule, Eq.\eqref{eq:argmax}, which assigns a class to all sources in the catalog, as well as the classification using the 1deg$^2$ threshold criteria and a very restrict threshold of 0.9.
Adding up all the objects predicted by the LFs results in $\sim 4,000$ objects with $r \in [17.5, 23.6]$.
However, the number of instances from the miniJPAS catalog within that interval is 7,468.
Therefore, we should not expect the numbers to agree perfectly with the LFs even when applying the argmax threshold.
On the other hand, the total number of objects using the threshold of 0.9 in CNN1 is 4,420.

\section{Conclusion}\label{conclusion}
In this work, which is part of the effort to identify quasars in the miniJPAS survey, we applied several machine learning models (CNN1, CNN2, LGBM and RF) to classify miniJPAS point-like sources as stars, galaxies, low-z (z < 2.1) and high-z (z $\geq$ 2.1) quasars, employing only photometry-based pseudo-spectra. 
In order to train and validate the models we used mock data catalogues developed by \citet{queiroz2022minijpas}.
The final miniJPAS quasar catalogue will be produced by combining the predictions from several classifiers (Pérez-Ràfols et al., in preparation), among them the ML models presented in this work, as well as those presented in Martínez-Solaeche et al. (in preparation) and Pérez-Ràfols et al. (in preparation).

In this paper we have constructed and tested five different ML models designed to be applied to miniJPAS data. 
We have focused on CNNs because of their potential to extract local features from the input data (the pseudo-spectra), and their ability to incorporate the information about errors in the data \citep{rodrigues2021information}.
We also applied well-established DT-based models as a baseline, and in order to complement the CNN approach.
We tested the robustness of the training sets by varying the populations of stars and retraining the models on modified samples, finding very small variations in the purity and completeness when training with these different data sets and applying to a fixed test set.
We have also checked, using permutation feature importance, that bluer filters are particularly relevant to correctly classifying high-z QSOs -- see Appendix \ref{fi}.

We evaluated the performance of the classifiers in terms of the purity and completeness of the predicted samples, and analysed the confusion between the four classes. We also investigated in more detail the sources of misclassification in terms of their luminosities, stellar types and redshifts. We verified that, as a general rule, the main source of confusion is between galaxies and low-z QSOs in the faint end. Stars are more often confused with low-z QSOs as well, specially bluer types (O, B, A, F), cataclysmic variables and white dwarfs. The redshift range of QSOs that were most often classified as galaxies is $z \in [1.0, 1.6]$.
The performances of the classifiers decrease as the objects become fainter and noisier. 
We verified that the predictions with a mock test set are indeed consistent with our previous knowledge about quasars, stars and galaxies features, which reinforces the quality of the mock data and also of the ML models developed in this work.

After validation, the ML models were finally applied to the miniJPAS data. For the few objects with spectroscopic confirmation of their classes, we obtained results consistent with the mock test sets (QSOs completeness $\sim 0.8$ and purity $\sim 0.95$).
Of the 7,468 point-like source in miniJPAS that lie in the magnitude range $17.5 < r \leq 23.6$, we found 2,309 stars, 3,827 galaxies, as well as 118 low-z QSOs and 547 high-z QSOs with CNN1 (noise model 11) and the 1deg$^2$ threshold criteria -- 667 objects did not have a type assigned with sufficient confidence to pass the thresholds specified in \S4.2.

When applying proper choices of probability thresholds to select the quasar candidates, the models underestimate the number of low-z QSOs and overestimate the number of high-z QSOs, specially in the very faint end, as compared to the LF from \citet{qsoLF_palanquedelabrouille}. 
Taken at face value, our results seem more consistent with the LF from \citet{qsoLF_croom2009}, which expects a higher number of faint high redshift QSOs as compared to \citet{qsoLF_palanquedelabrouille}.

This paper is another milestone in the J-PAS effort to map quasars at all redshifts with a minimal selection bias.
These quasars will be useful for a variety of applications: first, to study large-scale structure at high and intermediate redshifts, using both the QSOs themselves as tracers \citep{QSOLSS}, the Ly-$\alpha$ forest from their lines of sight \citep{LyaLSS}, that will be measured by the WEAVE instrument \citep{pieri2016weaveqso}, as well as their cross-correlations \citep{QSO-Lyacross}.
Second, to determine with higher accuracy both the quasar luminosity function \citep{qsoLF_croom2009,qsoLF_palanquedelabrouille} and the black holes mass function \citep{Jonas2022}, revealing the history of formation of those objects.
And finally, in the long run J-PAS should also be able to make a census of the QSOs including different sub-types that may be less represented in spectroscopic surveys due to the traditional targeting strategies.

\section*{Acknowledgements}
This paper has gone through internal review by the J-PAS collaboration.
N.R. acknowledges financial support from CAPES. R.A. was supported by CNPq and FAPESP.
C.Q. acknowledges financial support from FAPESP (grants 2015/11442-0 and 2019/06766-1) and Coordena\c{c}\~ao de Aperfei\c{c}oamento de Pessoal de N\'ivel Superior (Capes) -- Finance Code 001. I.P.R. was suported by funding from the European Union's Horizon 2020 research and innovation programme under the Marie Sklodowskja-Curie grant agreement No. 754510. M.P.P. and S.S.M. were supported by the Programme National de Cosmologie et Galaxies (PNCG) of CNRS/INSU with INP and IN2P3, co-funded by CEA and CNES, the A*MIDEX project (ANR-11-IDEX-0001-02) funded by the ``Investissements d'Avenir'' French Government program, managed by the French National Research Agency (ANR), and by ANR under contract ANR-14-ACHN-0021. G.M.S., R.M.G.D. and L.A.D.G. acknowledge support from the State Agency for Research of the Spanish MCIU through the ``Center of Excellence Severo Ochoa'' award to the Instituto de Astrof\'isica de Andaluc\'ia (SEV-2017-0709) and the project PID2019-109067-GB100. J.C.M. and S.B. acknowledge financial support from Spanish Ministry of Science, Innovation, and Universities through the project PGC2018-097585-B-C22. A.F.S. acknowledges support from the Spanish Ministerio de Ciencia e Innovaci\'on through project PID2019-109592GB-I00 and the Generalitat Valenciana project PROMETEO/2020/085. R.A.D. acknowledges partial support support from CNPq grant 308105/2018-4. AE acknowledges the financial support from the Spanish Ministry of Science and Innovation and the European Union - NextGenerationEU through the Recovery and Resilience Facility project ICTS-MRR-2021-03-CEFCA. LSJ acknowledges support from CNPq (304819/2017-4) and FAPESP (2019/10923-5). J.V. acknowledges the technical members of the UPAD for their invaluable work: Juan Castillo, Tamara Civera, Javier Hern\'andez, \'Angel L\'opez, Alberto Moreno, and David Muniesa. \\

Based on observations made with the JST250 telescope and PathFinder camera for the miniJPAS project at the Observatorio Astrof\'{\i}sico de Javalambre (OAJ), in Teruel, owned, managed, and operated by the Centro de Estudios de F\'{\i}sica del  Cosmos de Arag\'on (CEFCA). We acknowledge the OAJ Data Processing and Archiving Unit (UPAD) for reducing and calibrating the OAJ data used in this work. Funding for OAJ, UPAD, and CEFCA has been provided by the Governments of Spain and Arag\'on through the Fondo de Inversiones de Teruel; the Aragonese Government through the Research Groups E96, E103, E16\_17R, and E16\_20R; the Spanish Ministry of Science, Innovation and Universities (MCIU/AEI/FEDER, UE) with grant PGC2018-097585-B-C21; the Spanish Ministry of Economy and Competitiveness (MINECO/FEDER, UE) under AYA2015-66211-C2-1-P, AYA2015-66211-C2-2, AYA2012-30789, and ICTS-2009-14; and European FEDER funding (FCDD10-4E-867, FCDD13-4E-2685). Funding for the J-PAS Project has also been provided by the Brazilian agencies FINEP, FAPESP, FAPERJ and by the National Observatory of Brazil with additional funding provided by the Tartu Observatory and by the J-PAS Chinese Astronomical Consortium. \\
      
Funding for the Sloan Digital Sky Survey III/IV has been provided by the Alfred P. Sloan Foundation, the U.S. Department of Energy Office of Science, and the Participating Institutions. SDSS-III/IV acknowledge support and resources from the Center for High Performance Computing  at the University of Utah. The SDSS website is \url{www.sdss.org}. SDSS is managed by the Astrophysical Research Consortium for the Participating Institutions of the SDSS Collaboration including the Brazilian Participation Group, the Carnegie Institution for Science, Carnegie Mellon University, Center for Astrophysics | Harvard \& Smithsonian, the Chilean Participation Group, the French Participation Group, Instituto de Astrof\'isica de Canarias, The Johns Hopkins University, Kavli Institute for the Physics and Mathematics of the Universe (IPMU) / University of Tokyo, the Korean Participation Group, Lawrence Berkeley National Laboratory, Leibniz Institut f\"ur Astrophysik Potsdam (AIP), Max-Planck-Institut f\"ur Astronomie (MPIA Heidelberg), Max-Planck-Institut f\"ur Astrophysik (MPA Garching), Max-Planck-Institut f\"ur Extraterrestrische Physik (MPE), National Astronomical Observatories of China, New Mexico State University, New York University, University of Notre Dame, Observat\'ario Nacional / MCTI, The Ohio State University, Pennsylvania State University, Shanghai Astronomical Observatory, United Kingdom Participation Group, Universidad Nacional Aut\'onoma de M\'exico, University of Arizona, University of Colorado Boulder, University of Oxford, University of Portsmouth, University of Utah, University of Virginia, University of Washington, University of Wisconsin, Vanderbilt University, and Yale University. \\
      


\section*{Data Availability}

The final quasar catalogue will be generated with the combined code, described in the final article of the series (Pérez-Ràfols et al., in preparation).



\bibliographystyle{mnras}
\bibliography{main} 

\begin{thebibliography}{}
\makeatletter
\relax
\def\mn@urlcharsother{\let\do\@makeother \do\$\do\&\do\#\do\^\do\_\do\%\do\~}
\def\mn@doi{\begingroup\mn@urlcharsother \@ifnextchar [ {\mn@doi@}
  {\mn@doi@[]}}
\def\mn@doi@[#1]#2{\def\@tempa{#1}\ifx\@tempa\@empty \href
  {http://dx.doi.org/#2} {doi:#2}\else \href {http://dx.doi.org/#2} {#1}\fi
  \endgroup}
\def\mn@eprint#1#2{\mn@eprint@#1:#2::\@nil}
\def\mn@eprint@arXiv#1{\href {http://arxiv.org/abs/#1} {{\tt arXiv:#1}}}
\def\mn@eprint@dblp#1{\href {http://dblp.uni-trier.de/rec/bibtex/#1.xml}
  {dblp:#1}}
\def\mn@eprint@#1:#2:#3:#4\@nil{\def\@tempa {#1}\def\@tempb {#2}\def\@tempc
  {#3}\ifx \@tempc \@empty \let \@tempc \@tempb \let \@tempb \@tempa \fi \ifx
  \@tempb \@empty \def\@tempb {arXiv}\fi \@ifundefined
  {mn@eprint@\@tempb}{\@tempb:\@tempc}{\expandafter \expandafter \csname
  mn@eprint@\@tempb\endcsname \expandafter{\@tempc}}}

\bibitem[\protect\citeauthoryear{{Alam} et~al.,}{{Alam}
  et~al.}{2021}]{SDSS2021}
{Alam} S.,  et~al., 2021, \mn@doi [\prd] {10.1103/PhysRevD.103.083533}, \href
  {https://ui.adsabs.harvard.edu/abs/2021PhRvD.103h3533A} {103, 083533}

\bibitem[\protect\citeauthoryear{{Ata} et~al.,}{{Ata} et~al.}{2018}]{QSOLSS}
{Ata} M.,  et~al., 2018, \mn@doi [\mnras] {10.1093/mnras/stx2630}, \href
  {https://ui.adsabs.harvard.edu/abs/2018MNRAS.473.4773A} {473, 4773}

\bibitem[\protect\citeauthoryear{Baqui et~al.,}{Baqui et~al.}{2021}]{pbaqui}
Baqui P.~O.,  et~al., 2021, \mn@doi [Astronomy \& Astrophysics]
  {10.1051/0004-6361/202038986}, 645, A87

\bibitem[\protect\citeauthoryear{{Bautista} et~al.,}{{Bautista}
  et~al.}{2017}]{LyaLSS}
{Bautista} J.~E.,  et~al., 2017, \mn@doi [\aap] {10.1051/0004-6361/201730533},
  \href {https://ui.adsabs.harvard.edu/abs/2017A&A...603A..12B} {603, A12}

\bibitem[\protect\citeauthoryear{Benitez et~al.,}{Benitez
  et~al.}{2014}]{benitez2014jpas}
Benitez N.,  et~al., 2014, J-PAS: The Javalambre-Physics of the Accelerated
  Universe Astrophysical Survey (\mn@eprint {arXiv} {1403.5237})

\bibitem[\protect\citeauthoryear{{Blake} et~al.,}{{Blake}
  et~al.}{2012}]{WIGGLEZ}
{Blake} C.,  et~al., 2012, \mn@doi [\mnras] {10.1111/j.1365-2966.2012.21473.x},
  \href {https://ui.adsabs.harvard.edu/abs/2012MNRAS.425..405B} {425, 405}

\bibitem[\protect\citeauthoryear{Bonoli et~al.,}{Bonoli
  et~al.}{2021}]{miniJPAS}
Bonoli S.,  et~al., 2021, \mn@doi [Astronomy \& Astrophysics]
  {10.1051/0004-6361/202038841}, 653, A31

\bibitem[\protect\citeauthoryear{Breiman}{Breiman}{1996}]{breiman96_bagging}
Breiman L.,  1996, Machine Learning, 24, 123

\bibitem[\protect\citeauthoryear{Breiman}{Breiman}{2001}]{breiman2001rf}
Breiman L.,  2001, \mn@doi [Machine Learning] {10.1023/A:1010933404324}, 45, 5

\bibitem[\protect\citeauthoryear{Breiman, Friedman, Olshen  \& Stone}{Breiman
  et~al.}{1984}]{DT}
Breiman L.,  Friedman J.~H.,  Olshen R.~A.,   Stone C.~J.,  1984,
  Classification and Regression Trees.
Wadsworth International Group, Belmont, CA

\bibitem[\protect\citeauthoryear{{Burke}, {Aleo}, {Chen}, {Liu}, {Peterson},
  {Sembroski}  \& {Lin}}{{Burke} et~al.}{2019}]{2019MNRAS.490.3952B}
{Burke} C.~J.,  {Aleo} P.~D.,  {Chen} Y.-C.,  {Liu} X.,  {Peterson} J.~R.,
  {Sembroski} G.~H.,   {Lin} J. Y.-Y.,  2019, \mn@doi [\mnras]
  {10.1093/mnras/stz2845}, \href
  {https://ui.adsabs.harvard.edu/abs/2019MNRAS.490.3952B} {490, 3952}

\bibitem[\protect\citeauthoryear{Busca \& Balland}{Busca \&
  Balland}{2018}]{busca2018quasarnet}
Busca N.,  Balland C.,  2018, QuasarNET: Human-level spectral classification
  and redshifting with Deep Neural Networks (\mn@eprint {arXiv} {1808.09955})

\bibitem[\protect\citeauthoryear{Cabayol et~al.,}{Cabayol
  et~al.}{2018}]{cabayol2018}
Cabayol L.,  et~al., 2018, \mn@doi [Monthly Notices of the Royal Astronomical
  Society] {10.1093/mnras/sty3129}, 483, 529–539

\bibitem[\protect\citeauthoryear{{Chaves-Montero} et~al.,}{{Chaves-Montero}
  et~al.}{2017}]{eldar}
{Chaves-Montero} J.,  et~al., 2017, \mn@doi [\mnras] {10.1093/mnras/stx2054},
  \href {https://ui.adsabs.harvard.edu/abs/2017MNRAS.472.2085C} {472, 2085}

\bibitem[\protect\citeauthoryear{{Chaves-Montero} et~al.,}{{Chaves-Montero}
  et~al.}{2022}]{Jonas2022}
{Chaves-Montero} J.,  et~al., 2022, \mn@doi [\aap]
  {10.1051/0004-6361/202142567}, \href
  {https://ui.adsabs.harvard.edu/abs/2022A&A...660A..95C} {660, A95}

\bibitem[\protect\citeauthoryear{Chen \& Guestrin}{Chen \&
  Guestrin}{2016}]{xgboost}
Chen T.,  Guestrin C.,  2016, in Proceedings of the 22nd ACM SIGKDD
  International Conference on Knowledge Discovery and Data Mining. KDD '16.
ACM, New York, NY, USA, pp 785--794, \mn@doi{10.1145/2939672.2939785}, \url
  {http://doi.acm.org/10.1145/2939672.2939785}

\bibitem[\protect\citeauthoryear{Chollet et~al.}{Chollet
  et~al.}{2015}]{chollet2015keras}
Chollet F.,  et~al., 2015, Keras, \url {https://github.com/fchollet/keras}

\bibitem[\protect\citeauthoryear{{Cole} et~al.,}{{Cole} et~al.}{2005}]{2dF}
{Cole} S.,  et~al., 2005, \mn@doi [\mnras] {10.1111/j.1365-2966.2005.09318.x},
  \href {https://ui.adsabs.harvard.edu/abs/2005MNRAS.362..505C} {362, 505}

\bibitem[\protect\citeauthoryear{Cooper et~al.,}{Cooper
  et~al.}{2011}]{Cooper_2011}
Cooper M.~C.,  et~al., 2011, \mn@doi [The Astrophysical Journal Supplement
  Series] {10.1088/0067-0049/193/1/14}, 193, 14

\bibitem[\protect\citeauthoryear{{Croom} et~al.,}{{Croom}
  et~al.}{2009}]{qsoLF_croom2009}
{Croom} S.~M.,  et~al., 2009, \mn@doi [\mnras]
  {10.1111/j.1365-2966.2009.15398.x}, \href
  {https://ui.adsabs.harvard.edu/abs/2009MNRAS.399.1755C} {399, 1755}

\bibitem[\protect\citeauthoryear{{DES Collaboration} et~al.,}{{DES
  Collaboration} et~al.}{2021}]{DESCosmo2021}
{DES Collaboration} et~al., 2021, arXiv e-prints, \href
  {https://ui.adsabs.harvard.edu/abs/2021arXiv210513549D} {p. arXiv:2105.13549}

\bibitem[\protect\citeauthoryear{{Dalton}}{{Dalton}}{2016}]{dalton2016weave}
{Dalton} G.,  2016, in {Skillen} I.,  {Balcells} M.,   {Trager} S.,  eds,
  Astronomical Society of the Pacific Conference Series Vol. 507, Multi-Object
  Spectroscopy in the Next Decade: Big Questions, Large Surveys, and Wide
  Fields. p.~97

\bibitem[\protect\citeauthoryear{Davis et~al.,}{Davis
  et~al.}{2007}]{Davis_2007_aegis}
Davis M.,  et~al., 2007, \mn@doi [The Astrophysical Journal] {10.1086/517931},
  660, L1

\bibitem[\protect\citeauthoryear{Dawson et~al.,}{Dawson
  et~al.}{2013}]{Dawson_2012}
Dawson K.~S.,  et~al., 2013, \mn@doi [The Astronomical Journal]
  {10.1088/0004-6256/145/1/10}, 145, 10

\bibitem[\protect\citeauthoryear{{Dawson} et~al.,}{{Dawson}
  et~al.}{2016}]{Dawson2016AJ}
{Dawson} K.~S.,  et~al., 2016, \mn@doi [\aj] {10.3847/0004-6256/151/2/44},
  \href {https://ui.adsabs.harvard.edu/abs/2016AJ....151...44D} {151, 44}

\bibitem[\protect\citeauthoryear{Deng, Dong, Socher, Li, Li  \& Fei-Fei}{Deng
  et~al.}{2009}]{5206848}
Deng J.,  Dong W.,  Socher R.,  Li L.-J.,  Li K.,   Fei-Fei L.,  2009, in 2009
  IEEE Conference on Computer Vision and Pattern Recognition. pp 248--255,
  \mn@doi{10.1109/CVPR.2009.5206848}

\bibitem[\protect\citeauthoryear{{Dwelly} et~al.,}{{Dwelly}
  et~al.}{2017}]{Dwelly2017MNRAS}
{Dwelly} T.,  et~al., 2017, \mn@doi [\mnras] {10.1093/mnras/stx864}, \href
  {https://ui.adsabs.harvard.edu/abs/2017MNRAS.469.1065D} {469, 1065}

\bibitem[\protect\citeauthoryear{{Hikage} et~al.,}{{Hikage} et~al.}{2019}]{HSC}
{Hikage} C.,  et~al., 2019, \mn@doi [\pasj] {10.1093/pasj/psz010}, \href
  {https://ui.adsabs.harvard.edu/abs/2019PASJ...71...43H} {71, 43}

\bibitem[\protect\citeauthoryear{{Hoyle}, {Rau}, {Bonnett}, {Seitz}  \&
  {Weller}}{{Hoyle} et~al.}{2015}]{2015MNRAS.450..305H}
{Hoyle} B.,  {Rau} M.~M.,  {Bonnett} C.,  {Seitz} S.,   {Weller} J.,  2015,
  \mn@doi [\mnras] {10.1093/mnras/stv599}, \href
  {https://ui.adsabs.harvard.edu/abs/2015MNRAS.450..305H} {450, 305}

\bibitem[\protect\citeauthoryear{{Ivezi{\'c}} et~al.,}{{Ivezi{\'c}}
  et~al.}{2019}]{LSST_ivezic}
{Ivezi{\'c}} {\v{Z}}.,  et~al., 2019, \mn@doi [\apj]
  {10.3847/1538-4357/ab042c}, \href
  {https://ui.adsabs.harvard.edu/abs/2019ApJ...873..111I} {873, 111}

\bibitem[\protect\citeauthoryear{{Jansen} et~al.,}{{Jansen}
  et~al.}{2001}]{XMM_xray}
{Jansen} F.,  et~al., 2001, \mn@doi [\aap] {10.1051/0004-6361:20000036}, \href
  {https://ui.adsabs.harvard.edu/abs/2001A\&A...365L...1J} {365, L1}

\bibitem[\protect\citeauthoryear{Johnson \& Khoshgoftaar}{Johnson \&
  Khoshgoftaar}{2019}]{JohnsonK19}
Johnson J.~M.,  Khoshgoftaar T.~M.,  2019, J. Big Data, 6, 27

\bibitem[\protect\citeauthoryear{Ke, Meng, Finley, Wang, Chen, Ma, Ye  \&
  Liu}{Ke et~al.}{2017}]{lgbm}
Ke G.,  Meng Q.,  Finley T.,  Wang T.,  Chen W.,  Ma W.,  Ye Q.,   Liu T.-Y.,
  2017, Advances in neural information processing systems, 30, 3146

\bibitem[\protect\citeauthoryear{LeCun, Boser, Denker, Henderson, Howard,
  Hubbard  \& Jackel}{LeCun et~al.}{1989}]{CNN_lecun}
LeCun Y.,  Boser B.~E.,  Denker J.~S.,  Henderson D.,  Howard R.~E.,  Hubbard
  W.~E.,   Jackel L.~D.,  1989, Neural Computation, 1, 541

\bibitem[\protect\citeauthoryear{Lovell, Acquaviva, Thomas, Iyer, Gawiser  \&
  Wilkins}{Lovell et~al.}{2019}]{cnn_sfh}
Lovell C.~C.,  Acquaviva V.,  Thomas P.~A.,  Iyer K.~G.,  Gawiser E.,   Wilkins
  S.~M.,  2019, \mn@doi [Monthly Notices of the Royal Astronomical Society]
  {10.1093/mnras/stz2851}, 490, 5503

\bibitem[\protect\citeauthoryear{López-Sanjuan et~al.,}{López-Sanjuan
  et~al.}{2019}]{SGLC}
López-Sanjuan C.,  et~al., 2019, \mn@doi [Astronomy \& Astrophysics]
  {10.1051/0004-6361/201936405}, 631, A119

\bibitem[\protect\citeauthoryear{{Mar{\'\i}n-Franch}
  et~al.,}{{Mar{\'\i}n-Franch} et~al.}{2012}]{2012SPIE.8450E..3SM}
{Mar{\'\i}n-Franch} A.,  et~al., 2012, in {Navarro} R.,  {Cunningham} C.~R.,
  {Prieto} E.,  eds,  Society of Photo-Optical Instrumentation Engineers (SPIE)
  Conference Series Vol. 8450, Modern Technologies in Space- and Ground-based
  Telescopes and Instrumentation II. p. 84503S, \mn@doi{10.1117/12.925430}

\bibitem[\protect\citeauthoryear{Martí, Miquel, Castander, Gaztañaga, Eriksen
   \& Sánchez}{Martí et~al.}{2014}]{PAUS}
Martí P.,  Miquel R.,  Castander F.~J.,  Gaztañaga E.,  Eriksen M.,
  Sánchez C.,  2014, \mn@doi [Monthly Notices of the Royal Astronomical
  Society] {10.1093/mnras/stu801}, 442, 92

\bibitem[\protect\citeauthoryear{{Morganson} et~al.,}{{Morganson}
  et~al.}{2015}]{Morganson2015ApJ}
{Morganson} E.,  et~al., 2015, \mn@doi [\apj] {10.1088/0004-637X/806/2/244},
  \href {https://ui.adsabs.harvard.edu/abs/2015ApJ...806..244M} {806, 244}

\bibitem[\protect\citeauthoryear{{Myers} et~al.,}{{Myers}
  et~al.}{2015}]{Myers2015ApJS}
{Myers} A.~D.,  et~al., 2015, \mn@doi [\apjs] {10.1088/0067-0049/221/2/27},
  \href {https://ui.adsabs.harvard.edu/abs/2015ApJS..221...27M} {221, 27}

\bibitem[\protect\citeauthoryear{Nair \& Hinton}{Nair \&
  Hinton}{2010}]{Relubibcode}
Nair V.,  Hinton G.~E.,  2010, in Proceedings of the 27th International
  Conference on International Conference on Machine Learning. ICML'10.
Omnipress, Madison, WI, USA, pp 807--814

\bibitem[\protect\citeauthoryear{Nakazono et~al.,}{Nakazono
  et~al.}{2021}]{nakazono}
Nakazono L.,  et~al., 2021, \mn@doi [Monthly Notices of the Royal Astronomical
  Society] {10.1093/mnras/stab1835}, 507, 5847–5868

\bibitem[\protect\citeauthoryear{{Nakoneczny}, {Bilicki}, {Solarz}, {Pollo},
  {Maddox}, {Spiniello}, {Brescia}  \& {Napolitano}}{{Nakoneczny}
  et~al.}{2019}]{kids_qso_catalog}
{Nakoneczny} S.,  {Bilicki} M.,  {Solarz} A.,  {Pollo} A.,  {Maddox} N.,
  {Spiniello} C.,  {Brescia} M.,   {Napolitano} N.~R.,  2019, \mn@doi [\aap]
  {10.1051/0004-6361/201834794}, \href
  {https://ui.adsabs.harvard.edu/abs/2019A&A...624A..13N} {624, A13}

\bibitem[\protect\citeauthoryear{{Nakoneczny} et~al.,}{{Nakoneczny}
  et~al.}{2021}]{kids_ML_qso_selection}
{Nakoneczny} S.~J.,  et~al., 2021, \mn@doi [\aap]
  {10.1051/0004-6361/202039684}, \href
  {https://ui.adsabs.harvard.edu/abs/2021A&A...649A..81N} {649, A81}

\bibitem[\protect\citeauthoryear{Newman et~al.,}{Newman
  et~al.}{2013}]{Newman_2013}
Newman J.~A.,  et~al., 2013, \mn@doi [The Astrophysical Journal Supplement
  Series] {10.1088/0067-0049/208/1/5}, 208, 5

\bibitem[\protect\citeauthoryear{Palanque-Delabrouille
  et~al.,}{Palanque-Delabrouille et~al.}{2016}]{qsoLF_palanquedelabrouille}
Palanque-Delabrouille N.,  et~al., 2016, \mn@doi [Astronomy & Astrophysics]
  {10.1051/0004-6361/201527392}, 587, A41

\bibitem[\protect\citeauthoryear{{Pasquet}, {Bertin}, {Treyer}, {Arnouts}  \&
  {Fouchez}}{{Pasquet} et~al.}{2019}]{2019A&A...621A..26P}
{Pasquet} J.,  {Bertin} E.,  {Treyer} M.,  {Arnouts} S.,   {Fouchez} D.,  2019,
  \mn@doi [\aap] {10.1051/0004-6361/201833617}, \href
  {https://ui.adsabs.harvard.edu/abs/2019A&A...621A..26P} {621, A26}

\bibitem[\protect\citeauthoryear{Pedregosa et~al.,}{Pedregosa
  et~al.}{2011}]{scikit-learn}
Pedregosa F.,  et~al., 2011, Journal of machine learning research, 12, 2825

\bibitem[\protect\citeauthoryear{Pieri et~al.,}{Pieri
  et~al.}{2016}]{pieri2016weaveqso}
Pieri M.~M.,  et~al., 2016, WEAVE-QSO: A Massive Intergalactic Medium Survey
  for the William Herschel Telescope (\mn@eprint {arXiv} {1611.09388})

\bibitem[\protect\citeauthoryear{Pâris et~al.,}{Pâris
  et~al.}{2017}]{SDSSDR12}
Pâris I.,  et~al., 2017, \mn@doi [Astronomy \& Astrophysics]
  {10.1051/0004-6361/201527999}, 597, A79

\bibitem[\protect\citeauthoryear{Pérez-Ràfols, Pieri, Blomqvist, Morrison  \&
  Som}{Pérez-Ràfols et~al.}{2020}]{squeze}
Pérez-Ràfols I.,  Pieri M.~M.,  Blomqvist M.,  Morrison S.,   Som D.,  2020,
  \mn@doi [Monthly Notices of the Royal Astronomical Society]
  {10.1093/mnras/stz3467}, 496, 4931–4940

\bibitem[\protect\citeauthoryear{{Qu}, {Sako}, {M{\"o}ller}  \& {Doux}}{{Qu}
  et~al.}{2021}]{scone}
{Qu} H.,  {Sako} M.,  {M{\"o}ller} A.,   {Doux} C.,  2021, \mn@doi [\aj]
  {10.3847/1538-3881/ac0824}, \href
  {https://ui.adsabs.harvard.edu/abs/2021AJ....162...67Q} {162, 67}

\bibitem[\protect\citeauthoryear{Qu, Sako, Moller  \& Doux}{Qu
  et~al.}{2022}]{helenqu_22}
Qu H.,  Sako M.,  Moller A.,   Doux C.,  2022, A Convolutional Neural Network
  Approach to Supernova Time-Series Classification,
  \mn@doi{10.48550/ARXIV.2207.09440}, \url {https://arxiv.org/abs/2207.09440}

\bibitem[\protect\citeauthoryear{Queiroz et~al.,}{Queiroz
  et~al.}{2022}]{queiroz2022minijpas}
Queiroz C.,  et~al., 2022, The miniJPAS survey quasar selection I: Mock
  catalogues for classification (\mn@eprint {arXiv} {2202.00103})

\bibitem[\protect\citeauthoryear{{Ramachandra}, {Chaves-Montero}, {Alarcon},
  {Fadikar}, {Habib}  \& {Heitmann}}{{Ramachandra} et~al.}{2021}]{syth-z}
{Ramachandra} N.,  {Chaves-Montero} J.,  {Alarcon} A.,  {Fadikar} A.,  {Habib}
  S.,   {Heitmann} K.,  2021, arXiv e-prints, \href
  {https://ui.adsabs.harvard.edu/abs/2021arXiv211112118R} {p. arXiv:2111.12118}

\bibitem[\protect\citeauthoryear{Reis, Baron  \& Shahaf}{Reis
  et~al.}{2018}]{Reis_2018}
Reis I.,  Baron D.,   Shahaf S.,  2018, \mn@doi [The Astronomical Journal]
  {10.3847/1538-3881/aaf101}, 157, 16

\bibitem[\protect\citeauthoryear{Richards et~al.,}{Richards
  et~al.}{2002}]{Richards_2002}
Richards G.~T.,  et~al., 2002, \mn@doi [The Astronomical Journal]
  {10.1086/340187}, 123, 2945

\bibitem[\protect\citeauthoryear{{Richards} et~al.,}{{Richards}
  et~al.}{2009}]{2009ApJS..180...67R}
{Richards} G.~T.,  et~al., 2009, \mn@doi [\apjs] {10.1088/0067-0049/180/1/67},
  \href {https://ui.adsabs.harvard.edu/abs/2009ApJS..180...67R} {180, 67}

\bibitem[\protect\citeauthoryear{{Robin}, {Reyl{\'e}}, {Derri{\`e}re}  \&
  {Picaud}}{{Robin} et~al.}{2003}]{2003A&A...409..523R}
{Robin} A.~C.,  {Reyl{\'e}} C.,  {Derri{\`e}re} S.,   {Picaud} S.,  2003,
  \mn@doi [\aap] {10.1051/0004-6361:20031117}, \href
  {https://ui.adsabs.harvard.edu/abs/2003A&A...409..523R} {409, 523}

\bibitem[\protect\citeauthoryear{Rodrigues, Abramo  \& Hirata}{Rodrigues
  et~al.}{2021}]{rodrigues2021information}
Rodrigues N. V.~N.,  Abramo L.~R.,   Hirata N.~S.,  2021, The information of
  attribute uncertainties: what convolutional neural networks can learn about
  errors in input data (\mn@eprint {arXiv} {2108.04742})

\bibitem[\protect\citeauthoryear{Sharma, Kembhavi, Kembhavi, Sivarani, Abraham
  \& Vaghmare}{Sharma et~al.}{2019}]{cnn_stype}
Sharma K.,  Kembhavi A.,  Kembhavi A.,  Sivarani T.,  Abraham S.,   Vaghmare
  K.,  2019, \mn@doi [Monthly Notices of the Royal Astronomical Society]
  {10.1093/mnras/stz3100}, 491, 2280

\bibitem[\protect\citeauthoryear{{Sharma}, {Kembhavi}, {Kembhavi}, {Sivarani},
  {Abraham}  \& {Vaghmare}}{{Sharma} et~al.}{2020}]{2020MNRAS.491.2280S}
{Sharma} K.,  {Kembhavi} A.,  {Kembhavi} A.,  {Sivarani} T.,  {Abraham} S.,
  {Vaghmare} K.,  2020, \mn@doi [\mnras] {10.1093/mnras/stz3100}, \href
  {https://ui.adsabs.harvard.edu/abs/2020MNRAS.491.2280S} {491, 2280}

\bibitem[\protect\citeauthoryear{Shy, Tak, Feigelson, Timlin  \& Babu}{Shy
  et~al.}{2022}]{Shy_2022}
Shy S.,  Tak H.,  Feigelson E.~D.,  Timlin J.~D.,   Babu G.~J.,  2022, \mn@doi
  [The Astronomical Journal] {10.3847/1538-3881/ac6e64}, 164, 6

\bibitem[\protect\citeauthoryear{{Simonyan} \& {Zisserman}}{{Simonyan} \&
  {Zisserman}}{2014}]{2014arXiv1409.1556S}
{Simonyan} K.,  {Zisserman} A.,  2014, arXiv e-prints, \href
  {https://ui.adsabs.harvard.edu/abs/2014arXiv1409.1556S} {p. arXiv:1409.1556}

\bibitem[\protect\citeauthoryear{{Takada} et~al.,}{{Takada}
  et~al.}{2014}]{Takada2014PASJ}
{Takada} M.,  et~al., 2014, \mn@doi [\pasj] {10.1093/pasj/pst019}, \href
  {https://ui.adsabs.harvard.edu/abs/2014PASJ...66R...1T} {66, R1}

\bibitem[\protect\citeauthoryear{{Villacampa-Calvo}, {Zaldivar},
  {Garrido-Merch{\'a}n}  \& {Hern{\'a}ndez-Lobato}}{{Villacampa-Calvo}
  et~al.}{2021}]{2020arXiv200110523V}
{Villacampa-Calvo} C.,  {Zaldivar} B.,  {Garrido-Merch{\'a}n} E.~C.,
  {Hern{\'a}ndez-Lobato} D.,  2021, Journal of Machine Learning Research, \href
  {https://ui.adsabs.harvard.edu/abs/2020arXiv200110523V} {22, 1}

\bibitem[\protect\citeauthoryear{{Wolf}, {Meisenheimer}, {Rix}, {Borch}, {Dye}
  \& {Kleinheinrich}}{{Wolf} et~al.}{2003}]{Combo17}
{Wolf} C.,  {Meisenheimer} K.,  {Rix} H.~W.,  {Borch} A.,  {Dye} S.,
  {Kleinheinrich} M.,  2003, \mn@doi [\aap] {10.1051/0004-6361:20021513}, \href
  {https://ui.adsabs.harvard.edu/abs/2003A&A...401...73W} {401, 73}

\bibitem[\protect\citeauthoryear{{Wright} et~al.,}{{Wright}
  et~al.}{2010}]{WISE}
{Wright} E.~L.,  et~al., 2010, \mn@doi [\aj] {10.1088/0004-6256/140/6/1868},
  \href {https://ui.adsabs.harvard.edu/abs/2010AJ....140.1868W} {140, 1868}

\bibitem[\protect\citeauthoryear{{du Mas des Bourboux} et~al.,}{{du Mas des
  Bourboux} et~al.}{2017}]{QSO-Lyacross}
{du Mas des Bourboux} H.,  et~al., 2017, \mn@doi [\aap]
  {10.1051/0004-6361/201731731}, \href
  {https://ui.adsabs.harvard.edu/abs/2017A&A...608A.130D} {608, A130}

\makeatother
\end{thebibliography}




\clearpage

\appendix

\section{Additional Results} \label{app results}

\begin{figure*}
    \centering
    \includegraphics[width=1.0\linewidth]{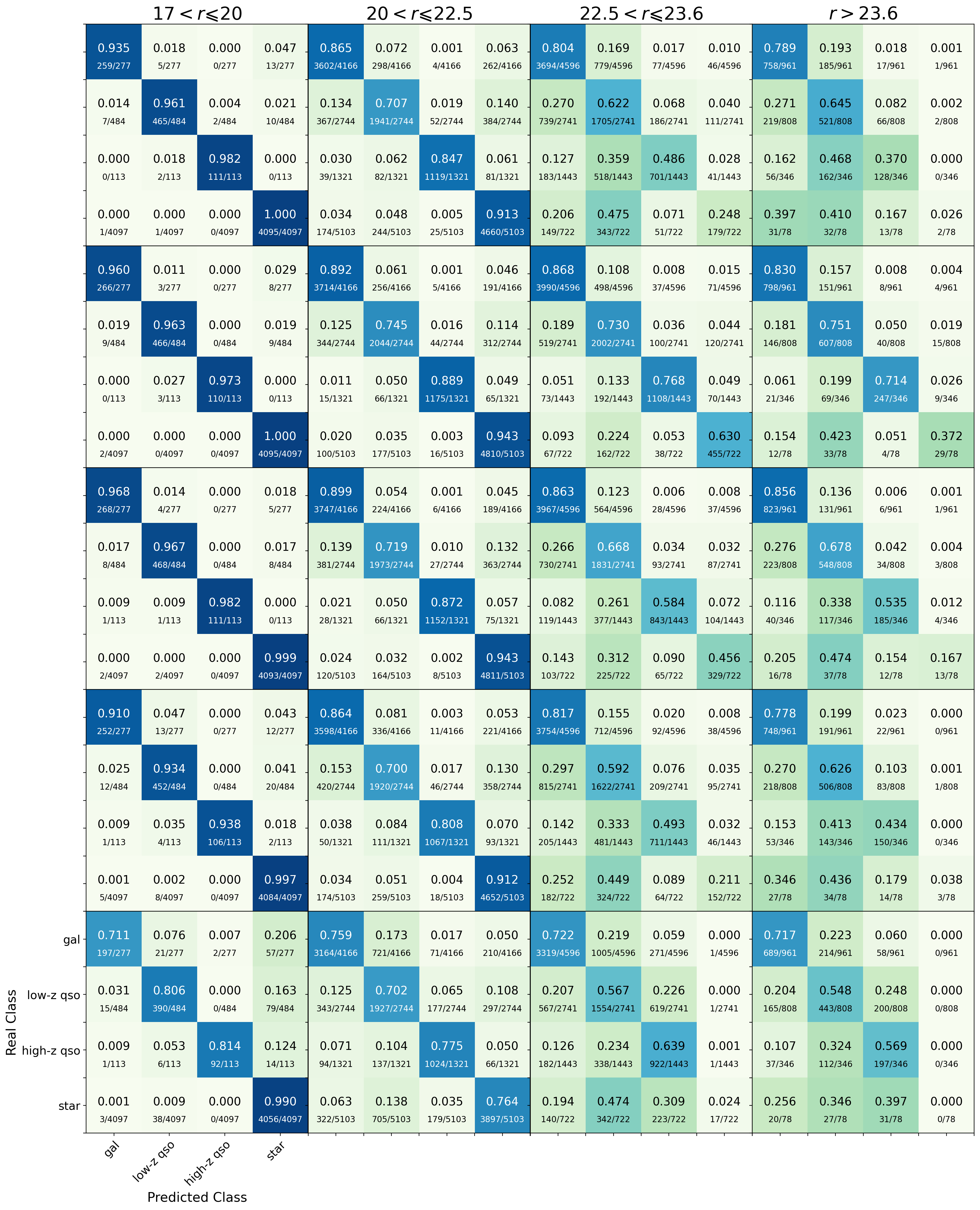}
    \caption{Confusion matrices of the classifiers in different $r$ band magnitude bins. From top to bottom: CNN1 without errors, CNN1, CNN2, LGBM, RF.}
    \label{fig:cm rbins}
\end{figure*}

Fig. \ref{fig:cm rbins} shows the confusion matrices in bins of $r$ magnitude. These results are summarised in the plots from Fig. \ref{fig:compare ML}.

\section{Feature Importance}\label{fi}
We performed a permutation feature importance analysis in the balanced test set to explore which features are more relevant for the models to make the predictions. We implemented this with the \texttt{eli5} package.
The procedure is the following: we exclude one filter at time and evaluate how the $F_1$ score of each class decreases with this missing filter. By ``exclusion'' of the filter we mean that the value of the filter becomes a random number, computed by combining the values of the features. Missing filters that lead to higher decrease in the performance are more important.

 \begin{figure*}
    \centering
    \includegraphics[width=0.9\linewidth]{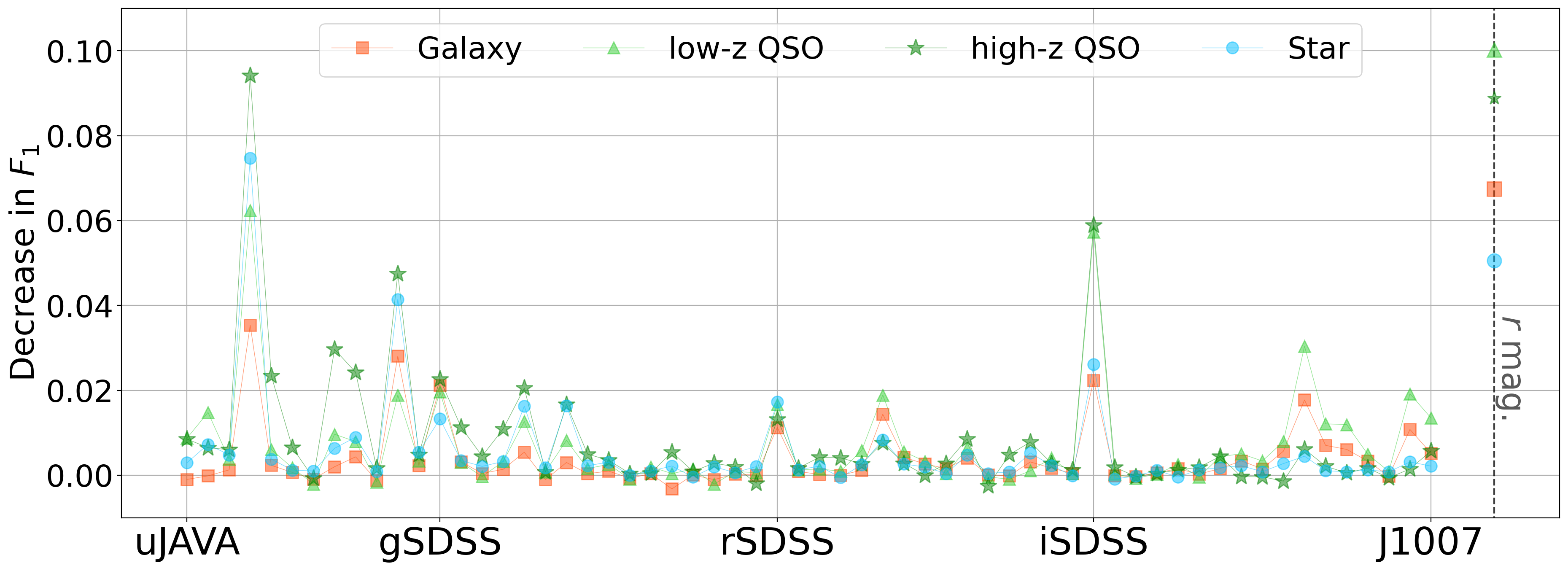}
    \caption{Permutation feature importance analysis in the balanced test set with LGBM. It computes the decrease in the $F_1$-score of each class when the measurement of a given filter is not available. The input features of LGBM (shown in the horizontal axis) are the normalised fluxes and the magnitude in the $r$ band (see \S\ref{ML DT}).}
    \label{fig:fi}
 \end{figure*}

Fig. \ref{fig:fi} shows the result of the permutation feature importance with LGBM in the balanced test set. We evaluate how much the $F_1$ score decreases as we remove each of the features.
We see that the exclusion of redder filters leads to a higher decrease in the $F_1$-score of low-z QSO, while for high-z QSO the bluer filters are more important.
The Ly-$\alpha$ and CIV emission lines are important features that characterize high-z QSOs. In the redshift range of $2.1 \leq z \leq 4.0$, the Ly-$\alpha$ line falls within $3,780\angstrom < \lambda < 6,080\angstrom$ and CIV falls within $4800\angstrom < \lambda < 7745\angstrom$, which could explain the importance of the filters that cover these wavelengths.

We re-trained LGBM excluding the 10 least important filters according to Fig. \ref{fig:fi} for each of the four classes. The results remained very similar, indicating that, although their contribution to the overall classification performance seems small, there is no clear advantage in excluding those features.


\section{CNN Settings}\label{cnn hp}
In this Section we describe the construction of CNN2 input data matrices (illustrated in Fig. \ref{fig:cnn2 input data}).
The parameters of the matrices are the number of columns, number of rows and the values to set the upper and lower boundaries (\texttt{n\_cols}, \texttt{n\_rows}, \texttt{up\_bound}, \texttt{low\_bound} -- see \citealt{rodrigues2021information}). A matrix is created by getting the mean value of the normalised fluxes (see Eq.\eqref{eq:norm flux}) of the object and by establishing the upper and lower values (the boundaries of the matrix) with \texttt{low\_bound} and \texttt{up\_bound}. In other words, if an object have the mean value equal to $\Bar{m}$, the matrix will cover the range of $[\Bar{m} - \texttt{low\_bound}$, $\Bar{m} + \texttt{up\_bound}]$.
The number of columns can be simply set as the number of attributes \texttt{n\_cols} = 60, since in our problem there is no uncertainty between the filters, i.e., a measurement certainly belongs to the given filter. The other parameters must be chosen more carefully to ensure that the matrix covers the complete J-spectra ranges and to ensure that the resolution of the pixels is large enough to properly resemble the probability distribution. Each filter have a specific probability distribution defined according to noise model 11. 
We set \texttt{n\_rows} = 90, \texttt{up\_bound} = 0.6 and \texttt{low\_bound} = 0.3. The resolution of the pixels is given by $(\texttt{up\_bound} + \texttt{low\_bound}) / \texttt{n\_rows} \approx 0.01$, which means that the probability distribution of the normalised fluxes is binned with intervals of $\approx 0.01$.

\section{Hyperparameter Tuning}\label{hp tune}

This Section describes the hyperparameter (HP) setting of the DT-based models. There are automated ways to set HPs, e.g, with grid search, but these might be computationally expensive. In this work, we performed a simple manual selection, by varying a few HPs that we consider relevant to monitor overfitting and underfitting. The best set of HPs was chosen according to the performance in the validation set. Parameters not shown were set as default.

For LGBM we tried varying the boosting type to search for better performance and computational gains, and the HPs shown in Table \ref{tab:lgbm hps} to monitor overfitting, such as the number of leaves and maximum depth of a tree. The number of trees (\texttt{n\_iterations}) is conditioned to \texttt{early\_stopping\_rounds}, which interrupts training after 200 iterations without improving the loss in the validation set.

For RF we tried different values for the parameters shown in Table \ref{tab:rf hps}. Although we do not limit the depth of each tree (\texttt{max\_depth} = None), we avoid overfitting by: (i) using the bagging strategy to create a tree, i.e., we set \texttt{bootstrap} = True, draw a sample (with replacement) equally sized to the training set (\texttt{max\_samples} = None) and sample $\sqrt{n}$ features (\texttt{max\_features} = `auto'), where $n$ is the total number of features; (ii) increasing the required number of instances to perform a split and to create a leaf in the trees (\texttt{min\_samples\_split}, \texttt{min\_samples\_leaf}, respectively). We also find an improvement by weighting the two types of quasars to match the proportion of stars and galaxies.
    
    \begin{table}
       \centering
       \caption{LGBM HP settings. Parameters not shown were set as default.}
       \begin{tabular}{ll}
        \hline
        hyperparameter & value \\
        \hline
        \texttt{objective} & `multiclass'\\
        \texttt{num\_class} & 4\\
        \texttt{boosting} & `gbdt'\\
        \texttt{learning\_rate} & 0.1\\
        \texttt{num\_leaves} & 31\\
        \texttt{max\_depth} & 6\\
        \texttt{early\_stopping\_rounds} & 200\\
        \hline
       \end{tabular}
       \label{tab:lgbm hps}
   \end{table}

    \begin{table}
       \centering
       \caption{RF HP settings. Parameters not shown were set as default.}
       \begin{tabular}{ll}
        \hline
        hyperparameter & value \\
        \hline
        \texttt{n\_estimators} & 100\\
        \texttt{criterion} & `gini'\\
        \texttt{max\_depth} & None\\
        \texttt{min\_samples\_split} & 100\\
        \texttt{min\_samples\_leaf} & 20\\
        \texttt{max\_features} & `auto'\\
        \texttt{max\_samples} & None\\
        \texttt{bootstrap} & True\\
        \texttt{random\_state} & 2\\
        \texttt{class\_weight} & \{0:1, 1:1, 2:1.47, 3:3.11\}\\
        \hline
       \end{tabular}
       \label{tab:rf hps}
   \end{table}


\bsp	
\label{lastpage}
\end{document}